\documentclass[aps,prb,twocolumn,superscriptaddress,floatfix]{revtex4}%
\usepackage{ams}
\usepackage{graphicx}
\usepackage{amsmath}
\usepackage{amsfonts}
\usepackage{amssymb}%
\def\nbZ{{\mathchoice {\hbox{$\sf\textstyle Z\kern-0.4em Z$}}
{\hbox{$\sf\textstyle Z\kern-0.4em Z$}} {\hbox{$\sf\scriptstyle
Z\kern-0.3em Z$}}  {\hbox{$\sf\scriptscriptstyle Z\kern-0.2em Z$}}}}
\begin{document}

\title{Protected Qubits and Chern Simons theories in Josephson Junction Arrays. }
\author{B. Dou\c{c}ot}
\affiliation{Laboratoire de Physique Th\'{e}orique et Hautes \'Energies, CNRS UMR 7589,
Universit\'{e}s Paris 6 et 7, 4, place Jussieu, 75252 Paris Cedex 05 France}
\author{M.V. Feigel'man}
\affiliation{Landau Institute for Theoretical Physics, Kosygina 2, Moscow, 117940 Russia}
\author{L.B. Ioffe}
\affiliation{Center for Materials Theory, Department of Physics and Astronomy, Rutgers
University 136 Frelinghuysen Rd, Piscataway NJ 08854 USA}
\altaffiliation{Landau Institute for Theoretical Physics, Kosygina 2, Moscow, 117940 Russia}

\author{A.S. Ioselevich}
\affiliation{Landau Institute for Theoretical Physics, Kosygina 2, Moscow, 117940 Russia}

\begin{abstract}
We present general symmetry arguments that show the appearance of doubly
denerate states protected from external perturbations in a wide class of
Hamiltonians. We construct the simplest spin Hamiltonian belonging to this
class and study its properties both analytically and numerically. We find that
this model generally has a number of low energy modes which might destroy the
protection in the thermodynamic limit. These modes are qualitatively different
from the usual gapless excitations as their number scales as the linear size
(instead of volume) of the system. We show that the Hamiltonians with this
symmetry can be physically implemented in Josephson junction arrays and that
in these arrays one can eliminate the low energy modes with a proper boundary
condition. We argue that these arrays provide fault tolerant quantum bits.
Further we show that the simplest spin model with this symmetry can be mapped
to a very special
${\mathchoice {\hbox{$\sf\textstyle Z\kern-0.4em Z$}} {\hbox{$\sf\textstyle
Z\kern-0.4em Z$}} {\hbox{$\sf\scriptstyle
Z\kern-0.3em Z$}} {\hbox{$\sf\scriptscriptstyle Z\kern-0.2em Z$}}}_{2}$
Chern-Simons model on the square lattice. We argue that appearance of the low
energy modes and the protected degeneracy is a natural property of lattice
Chern-Simons theories. Finally, we discuss a general formalism for the
construction of discrete Chern-Simons theories on a lattice.

\end{abstract}
\maketitle

\section{Introduction}

It is generally accepted that a quantum computer would have an enormous
advantage over the classical one for the solution of many fundamental and
practically important problems. \cite{Shor1994,Ekert1996,Steane1998} However,
its practical implementation presents a formidable challenge mostly because of
the conflicting requirements posed by scalability and decoupling from the
environment. In particular, all scalable designs are based on the solid state
devices but these are plagued by a strong decoherence. Quantitatively, in
physics it is conventional to measure the decoherence by the quality ratio,
$Q$, that is equal to the product of the decoherence time and a typical energy
gap while in computer science one uses the error rate, $R$, defined as the
probability of an error per time required for an individual operation. In order
to avoid excitations of higher energy states all operations should be performed
slowly on the scale of the inverse energy gap, so $R\gg1/Q$. The problem posed
by the omnipresence of the decoherence in solid state devices is exacerbated
by the fact that the error correction codes \cite{Shor1995}%
,\cite{Preskill1998} require a small error rate for individual qubits (at the
very least $10^{-3}$ per logical operation that translates into the quality
factors larger than $10^{4}$) and lead to a huge increase in the number of
qubits, effectively replacing one qubit by a lattice with $L\times L$ qubits
with large $L$ \cite{Kitaev2002}. Further, the efficient error correction
requires that operations are done \emph{simultaneously} on all $L\times L$
qubits encoding a single error free bit which makes this scheme rather
unrealistic. An alternative would be to use individual qubits with a very high
quality factor, much greater than $10^{8}$ which would allow to perform
calculations without (or with very little) error correction. If the noise
couples linearly to the energy difference between two states of the qubit
representing $0$ and $1$, this would mean that the physical noise should be
less than $10^{-8}$ times than all other energy scales of the device. It is
difficult to imagine a solid state physical system that is so well screened
from the outside noise, in particular, it is difficult to imagine a Josephson
junction, a Cooper box or SET where the motion of stray charges do not result
in a significant $1/f$ noise in electric or magnetic fields or in the strength
of the Josephson couplings. Further, a significant energy difference between
the two states of a qubit results in a phonon emission \cite{Ioffe2004} that 
limits the quality factor of a typical Josephson device by $10^{4}$. So,
the only logical possibility is that the physical noises do not affect the
energy difference between lowest energy states in the lowest orders (one or
more) in the strength of the noises. This is in principle possible because the
effect of the physical noise is always represented by the sum of local
physical operators (charge, current, etc). In the limiting case when the noise
does not affect the energy difference between the two states in any finite
order, these states form a protected subspace of the Hilbert space
\cite{Kitaev1997}. Of course, the formation of such truly protected space
becomes possible only for an infinite system. The implementation of these
models in solid state (Josephson junction) devices was suggested in
\cite{Ioffe2002a,Ioffe2002b,Doucot2003}. From a practical view point it is
important to consider simpler but smaller systems which are protected from the
noise in the given order, $n$, i.e. which are not affected by all physical
noises in all orders less than $n$. Clearly, any device in which two levels
representing $0$ and $1$ have a finite energy difference is susceptible to the
fluctuations in the physical quantity that sets this energy scale. For
instance, in a Cooper box this would be the Josephson, $E_{J}$, and a charging
energy, $E_{C}$ of the individual Josephson junction. Thus, even this limited
protection can occur only when the two "working" levels are degenerate.

It is well known that the stable degeneracy of the quantum levels is almost
always due to a high degree of the symmetry of the system. Examples are
numerous: time inversion invariance ensures the degeneracy of the states with
half-integer spin, rotational symmetry results in a degeneracy of the states
with non-zero momentum, etc. In order for the degeneracy to be stable with
respect to the local noise, one needs that the sufficient symmetry remains
even if a part of the system is excluded. The simplest example is provided by
the $6$ Josephson junctions connecting $4$ superconducting islands (so that
each island is connected with every other). \cite{Feigelman2003} In this
miniarray all islands are equivalent, so it is symmetric under all
transformations of the permutation group $S_{4}$. This group has a
two-dimensional representation and thus, pairs of the exactly degenerate
states. With the appropriate choice of the parameters one can make these
doublets the ground state of the system. The noise acting on one
superconducting island reduces the symmetry to the permutation group of the
three elements which still has two dimensional representations. So, this
system is protected from the noise in the first order ($n=2$). The goal of
this paper is to discuss designs giving the systems that are protected from
the noise in the higher orders. Note that systems with higher symmetry groups,
such as $5$ junctions connected by $10$ junctions (group $S_{5}$) typically do
not have two dimensional representations, so in these systems one can
typically get much higher degeneracy but not higher protection.

Generally, one gets degenerate states if there are two symmetry operations,
described by the operators $P$ and $Q$ that commute with the Hamiltonian but
do not commute with each other. If $[P,Q]|\Psi\rangle\neq0$ for any
$|\Psi\rangle$, all states are at least doubly degenerate. Local noise term is
equivalent to adding other terms in the Hamiltonian which might not commute
with these operators thereby lifting the degeneracy. Clearly, in order to
preserve the degeneracy one needs to have two \emph{sets (}of\emph{\ }$n$
elements each), $\{P_{i}\}$ and $\{Q_{i}\}$ of non-commuting operators, so
that any given local noise field does not affect some of them; further,
preferably, any given local noise should affect at most \emph{one} $P_{i}$ and
$Q_{i}$. In this case, the effect of the noise appears when $n$ noise fields
act \emph{simultaneously,} i.e. in the $n^{th}$ order in the noise strength.
Another important restriction comes from the condition that these symmetry
operators should not result in a higher degeneracy of the ideal system. For
two operators, $P$ and $Q$ that implies that $[P^{2},Q]=0$ and $[P,Q^{2}]=0$.
Indeed, one can construct the degenerate eigenstates of the Hamiltonian
starting with the eigenstate, $|0\rangle$, of $H$ and $Q$ and acting on this
state with $P.$ The resulting state, $|1\rangle$ should be different from the
original one because $P$ and $Q$ do not commute: $[P,Q]\Psi\neq0$ for any
$\Psi$. In a doubly degenerate system, acting again on this state with the
operator $P$ one should get back the state $|1\rangle$, so $[P^{2},Q]=0$. For
a set of operators, the same argument implies that in order to get a double
degeneracy (and not more) one needs that $[P_{i}P_{j},Q]=0$ and $[P,Q_{i}%
Q_{j}]=0$ for any $i,j$. Indeed, in this case one can diagonalize
simultaneously the set of operators $\{Q_{i}\}$, $\{Q_{i}Q_{j}\}$ and
$\{P_{i}P_{j}\}$. Consider a ground state, $|0\rangle$, of the Hamiltonian
which is also an eigenstate of all these operators. Acting on it with, say,
$P_{1}$ we get a new state, $|1\rangle$, but since $|1\rangle\propto
(P_{i}P_{1})P_{1}|0\rangle=P_{i}(P_{1}P_{1})|0\rangle\propto P_{i}|0\rangle$
all other operators of the same set would not produce a new state.

In the rest of the paper we introduce models that possess the symmetries
satisfying these conditions (Section II), discuss their possible
implementations in Josephson junction networks (Section III) and show that
they are equivalent to the Chern Simons ${\mathchoice {\hbox{$\sf\textstyle
Z\kern-0.4em Z$}} {\hbox{$\sf\textstyle Z\kern-0.4em Z$}} {\hbox{$\sf\scriptstyle
Z\kern-0.3em Z$}} {\hbox{$\sf\scriptscriptstyle Z\kern-0.2em Z$}}}_{2}$ gauge
theory (Sections IV and V). Section VI summarizes our results.

\section{Spin model}

The conditions discussed at the end of the previous section are fully
satisfied by the spin $S=1/2$ model on a square $n\times n$ array described by
the Hamiltonian
\begin{equation}
H=-J_{x}\sum_{i,j}\sigma_{i,j}^{x}\sigma_{i,j+1}^{x}-J_{z}\sum_{i,j}%
\sigma_{i,j}^{z}\sigma_{i+1,j}^{z}. \label{H}%
\end{equation}
Here $\sigma$ are Pauli matrices, note that the first term couples spins in
same row of the array while the second couples them along the columns. It is
not important for the following discussion whether the boundary conditions are
periodic or free, but since the latter are much easier to implement in a
hardware we shall assume them in the following. Further, the signs of the
couplings are irrelevant because for a square lattice one can always change it
by choosing a different spin basis on one sublattice. For the sake of
argument, we assumed that the signs of the couplings are ferromagnetic, this
is also a natural sign for Josephson junction implementations in Section III.
The Hamiltonian (\ref{H}) was first introduced in \cite{Kugel1982} as a model
for the anisotropic exchange interaction in transition metal compounds but its
properties remain largely unclear.

The Hamiltonian (\ref{H}) has two sets of the integrals of motion, $\{P_{i}\}
$ and $\{Q_{i}\}$ with $n$ operators each:
\begin{align*}
P_{i}  &  =%
{\displaystyle\prod\limits_{j}}
\sigma_{i,j}^{z}\\
Q_{j}  &  =%
{\displaystyle\prod\limits_{i}}
\sigma_{i,j}^{x}%
\end{align*}
i.e. each $P_{i}$ is the row product of $\sigma_{i,j}^{z}$ while $Q_{j}$ is
the column product of $\sigma_{i,j}^{x}$. Consider $P_{i}$ operator first. It
obviously commutes with the second term in the Hamiltonian and because the
first term contains two $\sigma_{i,j}^{z}$ operators in the same row, $P_{i}$
either contains none of them or both of them and since different Pauli
matrices anticommute, $P_{i}$ commutes with each term in the Hamiltonian
(\ref{H}). Similarly, $[Q_{i},H]=0$. Clearly, different $P_{i}$ commute with
themselves, $P_{i}^{2}=1$ and similarly $[Q_{i},Q_{j}]=0$ and $Q_{i}^{2}=1$,
but they do not commute with each other
\begin{align}
\{P_{i},Q_{j}\}  &  =0\label{[P,Q]}\\
\lbrack P_{i},Q_{j}]^{2}  &  =4\nonumber
\end{align}
so $[P_{i},Q_{j}]|\Psi\rangle\neq0$ for any $|\Psi\rangle$, thus in this model
all states are at least doubly degenerate. Further, because $P_{i}P_{j}$
contains two $\sigma_{i,j}^{z}$ in any column, such product commutes with all
$Q_{k}$ operators and similarly $[Q_{k}Q_{l},P_{i}]=0$. Thus, we conclude that
in this model all states are doubly degenerate, there is no symmetry reason
for larger degeneracy and that this degeneracy should be affected by the noise
only in the $n^{th}$ order of the perturbation theory.

To estimate the effect of the noise (which appears in this high order) one
needs to know the energy spectrum of the model and what are its low energy
states. All states of the system can be characterized by the set,
$\{\lambda_{i}=\pm1\}$ of the eigenvalues of $P_{i}$ operators (or
alternatively by the eigenvalues of the $Q_{j}$ operators). The degenerate
pairs of states are formed by two sets, $\{\lambda_{i}\}$ and $\{-\lambda
_{i}\}$ and each operator $Q_{j}$ interchanges these pairs: $Q_{i}%
\{\lambda_{i}\}=\{-\lambda_{i}\}$. We believe that different choices of
$\{\lambda_{i}=\pm1\}$ exhaust \emph{all} low energy states in this model,
i.e. that there are exactly $2^{n}$ low energy states. Note that this is a
somewhat unusual situation, normally one expects $n^{2}$ modes in a 2D system
and thus $2^{n^{2}}$ low energy states. The number $2^{n}$ low energy states
is natural for a one dimensional system and would also appear in two
dimensional systems if these states are associated with the edge. Here,
however, we can not associate them with the edge states because they do not
disappear for the periodic boundary conditions. We can not prove our
conjecture in a general case but we can see that it is true when one coupling
is much larger than the other and we have verified it numerically for the
couplings of the same order of magnitude. We start with the analytic treatment
of the $J_{z}\gg J_{x}$ case.

When one coupling is much larger than the others it is convenient to start
with the system where these others are absent and then treat them as small
perturbations; in the limit $J_{x}=0$ all columns are independent and the
ground state of each column is Ising ferromagnet. The ground state of each
column is doubly degenerate: $|1\rangle_{j}=%
{\displaystyle\prod\limits_{i}}
|\uparrow\rangle_{ij}$ and $|2\rangle_{j}=%
{\displaystyle\prod\limits_{i}}
|\downarrow\rangle_{ij} $ giving us $2^{n}$ degenerate states in this limit.
Excitations in each column are static kinks against the background of these
states, each kink has energy $2J_{z}$. Including now $J_{x}$ coupling, we see
that it creates two kinks in each of the neighboring columns increasing
thereby the energy of the system by $8J_{z}$ so the lowest order of the
perturbation theory is small in $J_{x}/8J_{z}$. The splitting between the
$2^{n}$ states occurs due to the high order processes which flip all spins in
two columns. In the leading approximation one can calculate the amplitude of
this process ignoring other columns. Thus, for this calculation we can
consider the model with only two columns that can be mapped onto a single
Ising chain in the transverse field in the following manner. The ground state
of two independent columns belongs to the sector of the Hilbert space
characterized by all $P_{i}P_{j}=1$, it is separated from the rest of the
spectrum by the gap of the order of $2J_{z}$. Further, the Hamiltonian does
not mix this sectors with different $P_{i}$, so in order to find the low
energy states, it is sufficient to diagonalize the problem in the sector
$P_{i}=1$ . In this sector only two states are allowed in each row:
$|\uparrow\uparrow\rangle$ and $|\downarrow\downarrow\rangle$, in the basis of
these states the Hamiltonian is reduced to
\begin{equation}
H_{\operatorname{col}}=-2J_{z}\sum_{i}\tau_{i}^{z}\tau_{i+1}^{z}-J_{x}\sum
_{i}\tau_{i}^{x} \label{H_col}%
\end{equation}
where $\tau$ are Pauli matrices acting in the space of $|\uparrow
\uparrow\rangle$ and $|\downarrow\downarrow\rangle$ states. This leads to the
splitting $2\Delta\approx(J_{x}/2J_{z})^{n}(2J_{z})$ \ (see Appendix A for the
details of this calculation)\ between the symmetric and antisymmetric
combinations of the two ferromagnetic chains in this problem. Thus, we
conclude that the effective Hamiltonian of the low energy states in the full
system is%

\[
H=\Delta\sum_{j}\widehat{\tau}_{j}^{x}\widehat{\tau}_{j+1}^{x}
\]
where $\widehat{\tau}$ are Pauli matrices acting in the space of $|1\rangle$
and $|2\rangle$ states describing the global state of the whole column. This
effective low energy model also describes a ferromagnetic chain in which the
excitations (static kinks) are separated from the degenerate ground state by
the gap $2\Delta$. In the basis of these $2^{n}$ low energy states the
operators $Q_{j}=\widehat{\tau}_{j}^{x}.$

We conclude that \ in the limit $J_{z}\gg J_{x}$ $2^{n}$ low energy states
form a narrow (of the order of $\Delta$) band inside a much larger gap,
$J_{z}$, characterized by different eigenvalues of $Q_{j}$ operators and by
one value $P_{i}P_{j}=1$. In the opposite limit $J_{x}\gg J_{z}$ low energy
states form a narrow band inside the gap of size $J_{x}$ characterized by
different eigenvalues of $P_{j}$ operators and by the same value $Q_{i}%
Q_{j}=1$. Consider now the effect of a weak noise in the former limit. To be
more specific we consider the effect of the additional single site fields:
\[
H_{n}=\sum_{i,j}h_{i,j}^{z}\sigma_{i,j}^{z}+h_{i,j}^{x}\sigma_{i,j}^{x}
\]
The first term shifts (up or down) the energies of each ferromagnetic column
by $H_{i}^{z}=\sum_{i}h_{i,j}^{z}$ while the second term gives the transitions
between up and down states in each column. These transitions appear only in
the order $n$ of the perturbation theory, so their amplitude is exponentially
small: $H_{j}^{x}=(%
{\displaystyle\prod\limits_{i}}
h_{ij}^{x}/J_{z})J_{z}$. Thus, when projected onto the low energy subspace
this noise part becomes
\[
H_{n}=\sum_{j}H_{j}^{z}\widehat{\tau}_{j}^{z}+H_{j}^{x}\widehat{\tau}_{j}^{x}
\]
The effect of the first term on the ground state degeneracy appears in the
$n^{th}$ order of the perturbation theory in $H_{j}^{z}/\Delta$ and so it is
much bigger than that of the second term because $\Delta$ becomes
exponentially small as $n\rightarrow\infty$ for $J_{x}\ll J_{z}$. Note that
although the effect of the $h_{i,j}^{z}\sigma_{i,j}^{z}$ noise appears only in
the large order of the pertubation theory, it is not small because of the
small energy denominator in this parameter range. Similarly, we expect that in
the opposite limit, $J_{z}\ll J_{x}$, the low lying states are characterized
by the set of eigenvalues of $P_{i}$ operators, the effect of the $h_{i,j}%
^{x}\sigma_{i,j}^{x}$ grows rapidly while the effect of the $h_{i,j}^{z}%
\sigma_{i,j}^{z}$ noise decreases with the $J_{x}$ increase. We conclude that
in the limits when one coupling is much larger than another ($J_{z}\gg J_{x}$
or $J_{z}\ll J_{x}$), the gap closes very quickly (exponentially) and the
non-linear effect of the appropriate noise grows rapidly with the system size.
These qualitative conclusions should remain valid for all couplings except a
special isotropic ($J_{x}=J_{y}$ ) point unless the system undergoes a phase
transition near this point (at some $J_{x}/J_{y}=j_{c}\sim1$).

\begin{figure}[th]
\includegraphics[width=3.0in]{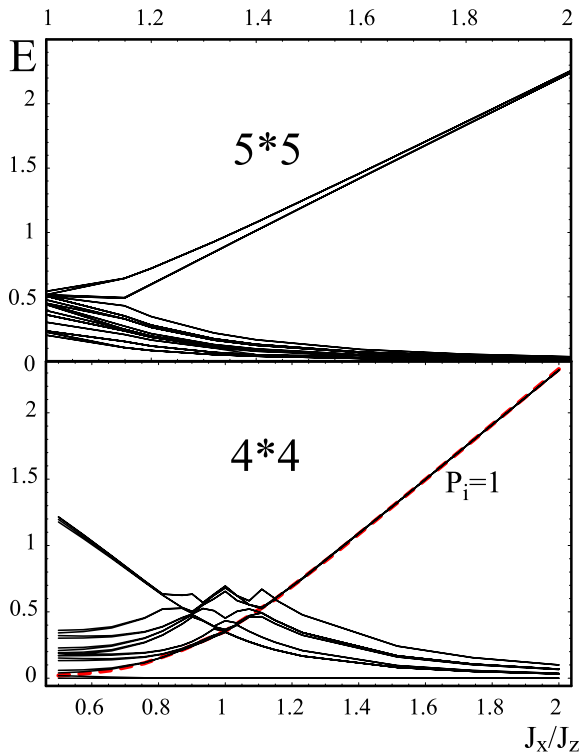}\caption{Energy spectrum of the
$5\times5$ and $4\times4$ systems in the units of $J_{z}$ coupling as a
function of $J_{x}/J_{z}$. We show energies of the lowest 40 states for
$5\times5$ (upper pane) and lowest 20 states for the $4\times4$ system (lower
pane). One clearly sees that at large anisotropy a well defined low energy
band is formed which contains $2^{5}$ states for $5\times5$ system and $2^{4}$
states for $4\times4$ one. In order to verify that low energy states are in
one-to-one correspondence with $P_{i}$ eigenvalues for large $J_{x}/J_{z}$ we
have calculated the second lowest eigenstate in $P_{i}=1$ sector (first one is
the ground state). As shown in the lower pane, this state indeed has a large
gap for $J_{x}/J_{z}\gtrsim1.2$ }%
\label{Spectra}%
\end{figure}

In order to check these conclusions we have numerically diagonalized small
spin systems containing up to 5 by 5 spins subjected to a small random field
$h_{i,j}^{z}$ flatly distributed in the interval $(-\delta/2,\delta/2)$. We
see that indeed the gap closes rather fast away from the special $J_{x}=J_{y}$
point (Fig.~\ref{Spectra}) but remains significant near $J_{z}=J_{x}$ point
where it clearly has a much weaker size dependence. Interestingly, the gap
between the lowest $2^{n}$ states and the rest of the spectrum expected in the
limits $J_{z}\gg J_{x}$ or $J_{z}\ll J_{x}$ appears only at $J_{x}/J_{z}%
>j_{c}$ with a practically size independent $j_{c}\approx1.2$. We also see
that the condition $P_{i}=1$ eliminates all low lying states in the $J_{z}\ll
J_{x}$ limit where the lowest excited state in $P_{i}=1$ sector is separated
from the ground state by a large gap and in fact provides a lower bound for
all high energy states. The special nature of this state appears only at
$J_{x}/J_{z}>j_{c}$. Clearly, the system behaves quite differently near the
isotropic point and away from it but the size limitations do not allow us to
conclude whether these different regimes correspond to two different phases
(with the \textquotedblright isotropic\textquotedblright\ phase restricted to
a small range of parameters $j_{c}^{-1}<J_{x}/J_{z}<j_{c}$) or it is a
signature of the critical region which becomes narrower as the size increases.
Although we do not see any appreciable change in $j_{c}$ with the system size,
our numerical data do not allow us to exclude the possibility that $j_{c}$
tends to unity in the thermodynamic limit. We conclude that numerical data
\textit{favor} intermediate phase scenario. In contrast to this, the
analytical resuls for two and three chains indicate that the transition occurs
only at $J_{z}=J_{x} $ point. Namely, both two and three chain models with
periodic boundary conditions in the transverse direction can be mapped onto
solvable models with transition at $J_{z}=J_{x}$: in the case of two chains
the problem is mapped onto the exactly solvable Ising model in transverse
field as described above while the three chain model is mapped (see Appendix
B) onto the 4 states Potts model in a similar way. The latter is not exactly
solvable in the whole parameter range but it obeys the exact duality that
allows one to determine its critical point \cite{Baxter,Wu} , further, its
exponents can be determined from the conformal field theory
\cite{PottsExponents}. The mapping of three chain problem onto the Potts model
is possible because the number of states of three spin rung for a given value
of the conserved $P$ operator is $4$ while the number of different terms in
the Hamiltonian that couples the adjacent rungs is $3$. For a larger number of
chains the number of states in each rung grows exponentially while the number
of terms in the Hamiltonian grows only linearly making such mappings
impossible. In this sense two and three chain models are exceptional and it is
fairly possible that the intermediate phase appears in models with larger
number of chains.

\begin{figure}[th]
\includegraphics[width=3.0in]{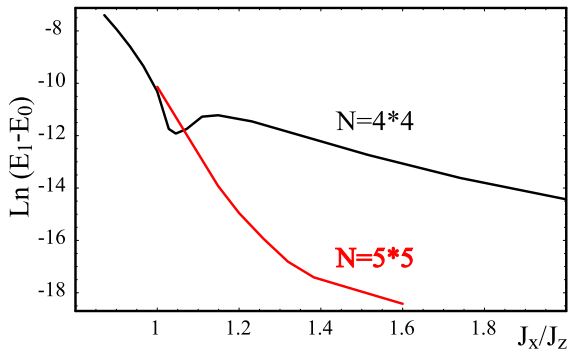}\caption{Ground state splitting by
random field in $z$-direction for $5\times5$ and $4\times4$ systems. The
random field acted on each spin and was randomly distributed in the interval
$(-0.05,0.05)$. Note that the effect of the random field in z-direction
becomes larger for $J_{x}\ll J_{z}$ as expected (see text). Because near
$J_{x}=J_{z}$ isotropic point the gap for $5\times5$ system is significantly
smaller than the gap for the $4\times4$ system, this relatively large disorder
has almost the same effect on these systems at $J_{x} \sim J_{z}$. We have
verified numerically that decrease of the disorder by a factor of two leads to
a dramatically smaller effects for $5\times5$ system confirming the scaling
$E_{1}-E_{0}\propto\delta^{n}$ discussed in the text; for $J_{x}/J_{z}>1.2$
the difference $E_{1}-E_{0}$ is difficult to resolve numerically.}%
\label{Splitting}%
\end{figure}

Finally, we checked the effect of the $h_{i,j}^{z}\sigma_{i,j}^{z}$ on the
ground state degeneracy splitting, our results are shown in
Fig.~\ref{Splitting}. We see that, as expected, this disorder becomes relevant
for $J_{x}\ll J_{z}$ while in \ the opposite limit its effect quickly becomes
unobservable. We conclude that at (and perhaps near) isotropic point, the gap
closes slowly with the system size and the effect of even significant disorder
($\delta=0.1$) becomes extremely small for the medium sized systems.

Although it is not clear how fast the gap closes in thermodynamic limit (if it
closes exponentially fast the system never becomes truly protected from the
noise because the effect of the high order terms might get very large), our
numerical results clearly indicate that medium size (4 by 4 or 5 by
5)\ systems provide an extremely good protection from the noise suppressing
its effect by many orders of magnitude. This should be enough for all
practical purposes. Further, if it is possible to construct the systems where
$P_{i}P_{j}=1$ (in other words with an additional term in the Hamiltonian
$H_{P}=-\Delta\sum_{ij}P_{i}P_{j}$ with significant $\Delta$), this would
eliminate the dangerous low energy states, leading to a good protection for
all couplings strengths $J_{x}\geq J_{z}$. Indeed, in this case, we can repeat
the previous analysis and conclude that the effects of the noise appear only
in the $n^{th}$ order and that now the perturbation theory in the
\textquotedblright dangerous\textquotedblright\ noise, $H_{j}^{z}$, implies
the expansion in $H_{j}^{z}/\Delta$ where $\Delta$ is no longer exponentially
small but is the coefficient in the Hamiltonian\ $H_{P}$. Thus, in this case
these higher order terms become small.

\section{Josephson junction implementations}

\bigskip The basic ingredient of any spin $1/2$ implementation in Josephson
junction array is the elementary block that has two (nearly degenerate)
states. One of the simplest implementation is provided by a four Josephson
junction loop (shown as rhombus in Fig. \ref{SArray}a) penetrated by magnetic
flux $\Phi_{0}/2$. \cite{Blatter2001,Doucot2002}. Classically, this loop is
frustrated and its ground state is degenerate: it corresponds to the phase
differences $\pm\pi/4$ across each junction constituting the loop. Two states
(spin \textquotedblright up\textquotedblright\ and \textquotedblright
down\textquotedblright) then correspond to the states with phase difference
$\pm\pi/2$ across the rhombus. For a isolated rhombus a non-zero (but
small)\ charging energy, $E_{C}=e^{2}/2C$, would result in the transitions
between these two states with the amplitude
\begin{equation}
r\approx E_{J}^{3/4}E_{C}^{1/4}e^{-s\sqrt{E_{J}/E_{C}}} \label{r}%
\end{equation}
thereby lifting this degeneracy. Here $s$ is numerical coefficient of the
order of unity that was found in \cite{Ioffe2002b}, $s\approx1.61$ and $E_{J}$
is the Josephson energy of each junction.

\subsection{Simplest Josephson junction array.}

\begin{figure}[th]
\includegraphics[width=3.0in]{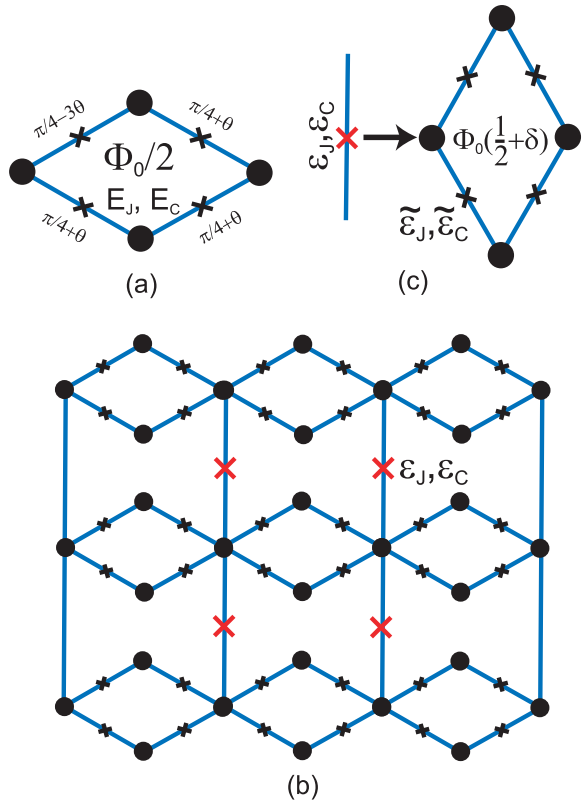}\caption{ Schematics of the array
equivalent to the spin model with the interaction (\ref{H_z}) in the vertical
direction. (a) the main element of the array, the superconducting rhombus
frustrated by magnetic flux $\Phi_{0}/2$. Josephson energy of each rhombus is
minimal for $\theta=0$ and $\theta=-\pi/2$. Significant charging energy
induces the transitions $\theta=0 \leftrightarrow\theta=-\pi/2$ between these
energy minima. (b) The array geometry. The superconducting boundary conditions
chosen here ensure that $P_{i} P_{j}=1$ thereby eliminating all low lying
states in the appropriate regime (see text). (c) The requirement that
continuous phase does not fluctuate much while the discrete variables have
large fluctuations is easier to satisfy in a very big arrays ($L>20$) if one
replaces the vertical links by the rhombi with junctions with $\tilde
{\epsilon_{J}}, \tilde{\epsilon_{C}}$ frustrated by the flux $\Phi_{0}%
(\delta+1/2)$ }%
\label{SArray}%
\end{figure}

\bigskip We begin with the Josephson junction array that has two sets of the
integrals of motion, $\{P_{i}\}$ and $\{Q_{i}\}$ discussed above, which is
shown in Fig.~\ref{SArray}. This array contains rhombi with junctions
characterized by Josephson and charging energies $E_{J}\gtrsim E_{C}$ and a
weaker vertical junctions characterized by the energies $\epsilon_{J}%
,\epsilon_{C}$. As we explain below, although this array preserves the
integrals of motion, $\{P_{i}\}$ and $\{Q_{i}\}$, it maps onto a spin model
that differs from (\ref{H}); we consider a more complicated arrays that are
completely equivalent to spin model (\ref{H}) in the next subsection. The
state of the system is fully characterized by the state of each rhombi
(described by the effective spin $1/2$) and by the small deviations of the
continuous superconducting phase across each junction from its equilibrium
(classical) values. Ignoring for the moment the continuous phase, we see that
the potential energy of the array shown in Fig. \ref{SArray} is given by
\begin{equation}
H_{z}=-\epsilon_{J}\sum_{i,j}\tau_{i,j}^{z}\tau_{i+1,j}^{z},\qquad\tau
_{i,j}^{z}\equiv\prod_{k<j}\sigma_{i,k}^{z} \label{H_z}%
\end{equation}
Physically, the variable $\tau_{i,j}^{z}$ describes the phase of the rightmost
corner of each rhombi with respect to the left (grounded)\ superconducting
wire modulo $\pi$. The right superconducting wire (that connects the rightmost
corners of the rhombi in the last column) ensures that the phase differences
along all rows are equal. In the limit of a large phase stiffness this implies
that the number of the rhombi with the phase difference $\pi/2$ should be
equal for all rows modulus 2. This constraint does not allow an individual
rhombus flips, instead a flip of one rhombus should be always followed by a
flip of another in the same row. If, however, the phase stiffness is low, the
flip of one rhombus can be also compensated by the continous phase
deformations in the other rhombi constituting this row, we derive the
conditions at which we can exclude these processes below. The simplest allowed
process is the simultaneous flip of two rhombi in one row%
\begin{equation}
H_{x}=-\sum_{i,j,k}\widetilde{J_{x}}(j-k)\sigma_{i,j}^{x}\sigma_{i,k}^{x}
\label{H_x_lr}%
\end{equation}
where $\widetilde{J_{x}}(k)$ is the amplitude to flip two rhombi a distance
$k$ apart. Both potential (\ref{H_z}) and kinetic (\ref{H_x_lr}) energies
commute with the integrals of motion, $\{P_{i}\}$ and $\{Q_{j}\}$, so that we
expect that the main feature of this model, namely, the existence of the
protected doublets will be preserved by this array.

\bigskip As explained in the previous section, in order to achieve a really
good protection one needs to eliminate all low energy states (except for the
degenerate ground state) characterized by different values of the $\{P_{i}\}$
and $\{Q_{j}\}$ operators. The array shown in Fig. \ref{SArray} has a boundary
conditions implying $P_{i}P_{j}=1$ for any $i,j$ because in this array the sum
along each row of the phases across individual rhombus should be equal for all
rows. Thus, for a sufficiently large tunneling amplitude $\widetilde{J_{x}%
}(k)$ this array should have two degenerate ground states separated from the
rest of the spectrum by a large gap. Physically, these two states correspond
to two different values of the phase difference along each row. The
quantitative condition ensuring that tunneling amplitude $\widetilde{J_{x}%
}(k)$ is large enough depends on the range of $\widetilde{J_{x}}(k)$. The
simplest situation is realized if only the nearest neighboring rhombi flip
with the significant amplitude, $J_{x}$. Because flip of the two nearest
rhombi is equivalent to the flip of the phase on the island between them, in
this case the spin model (\ref{H_z},\ref{H_x_lr}) is equivalent to the
collection of independent vertical Ising chains with Hamiltonian%
\[
H=-\sum_{i,j}\epsilon_{J}\tau_{i,j}^{z}\tau_{i+1,j}^{z}-J_{x}\tau_{i,j}^{x}
\]
For $J_{x} \gg2\epsilon_{J}$ each chain described by this Hamiltonian has a
unique ground state separated by the $\Delta=2J_{x}$ from the rest of the
spectrum. As the ratio $\epsilon_{J}/J_{x}$ grows, the gap decreases.

In the opposite limiting case of a very long range $\widetilde{J_{x}}%
(k)=J_{x}$, one can treat the interaction (\ref{H_x_lr}) in the mean field
approximation%
\begin{equation}
H_{x}=-J_{x}L_{x}\langle\sigma_{i,k}^{x}\rangle\sum_{i,j}\sigma_{i,j}^{x}%
\end{equation}
At large $J_{x}$ the ground state of this system is also a doublet
(characterized by $\langle\sigma_{i,k}^{x}\rangle=\pm1$) with all other
excitations separated by the gap $\Delta=2L_{x}J_{x}$ from the rest of the
spectrum. As we increase the vertical coupling, $\epsilon_{J}$, the gap for
the excitations gets smaller. At very large $\epsilon_{J}$ the Hamiltonian is
dominated by the ferromagnetic coupling in the vertical direction, so in this
regime there are many low energy states corresponding to two possible
magnetizations of each column. The magnitude of $\epsilon_{J}$ for which the
gap decreases significantly can be estimated from the first order correction
in $\epsilon_{J}$. The dominant contribution comes from the transitions
involving rhombi of the outmost rows. They occur with amplitude $\epsilon_{J}$
and lead to the states with energy $\Delta$, so we expect that as long as
$\epsilon_{J}\lesssim\Delta$, the system has a doubly degenerate ground state
separated from the other states by gap of the order of $\Delta$.

The amplitude, $\widetilde{J_{x}}(k)$, for the simultaneous flip of two rhombi
can be found from the same calculation that was used in \cite{Ioffe2002b} to
calculate a single rhombus flip and the simultaneous flip of three rhombi. If
$\epsilon_{C}\gg E_{C}$, the contribution of the vertical links to the total
kinetic energy of the superconducting phase is small and can be treated as a
small perturbation, in this case
\begin{equation}
\widetilde{J_{x}}(k)\approx E_{J}^{3/4}E_{C}^{1/4}e^{-2s\sqrt{E_{J}/E_{C}%
}(1+ck\frac{E_{C}}{\epsilon_{C}})} \label{J_x(k)}%
\end{equation}
where $c\sim1$. Here the factor 2 in the exponential appears because in this
process one changes simultaneously the phases across two neighboring rhombi.
Note that although the relative change in the action due to vertical links is
always small, their contribution might suppress the flips of all rhombi except
the nearest neighbors if $E_{J}E_{C}/\epsilon_{C}^{2}\gg1$. Note that even a
relatively large $\epsilon_{C}$ (so that $E_{C}/\epsilon_{C}\ll1$) can be
sufficient to suppress the processes involving non-nearest neighbors. We
conclude that the low energy states become absent as long as
\begin{equation}
\epsilon_{J}<L_{eff}J_{x} \label{epsilon_J}%
\end{equation}
where $L_{eff}=1/2$ if $E_{J}E_{C}/\epsilon_{C}^{2}\gg1$ and $L_{eff}%
\approx\min(\epsilon_{C}/\sqrt{E_{J}E_{C}},L)$ if $E_{J}E_{C}/\epsilon_{C}%
^{2}\ll1$. These estimates assume that the main contribution to the
capacitance comes from the junctions and ignores the contribution from the
self-capaciatnce. If the self-capacitance is significant, the processes
involving more than one island become quickly suppressed.

We now consider the effect of the continuous fluctuations of the
superconducting phase. Generally, a finite phase rigidity allows single
rhombus flip, described by the $\sum_{ij}\widetilde{t}\sigma_{ij}^{x}$ term in
the effective spin Hamiltonian. This term does not commute with the intergrals
of motion $P_{i}$ and thereby destroys the protected doublets. However, for a
significant phase rigidity the energy of a state formed by a single rhombus
flip, $U_{sf}$, is large. If, further, the amplitude $\widetilde{t}$ of these
processes is small: $\widetilde{t}\ll U_{sf}$, the states corresponding to
single flips can be eliminated from the effective low energy theory and the
protection is restored. If $\widetilde{t}>U_{sf}$, the protection is lost.

\begin{figure}[th]
\includegraphics[width=3.0in]{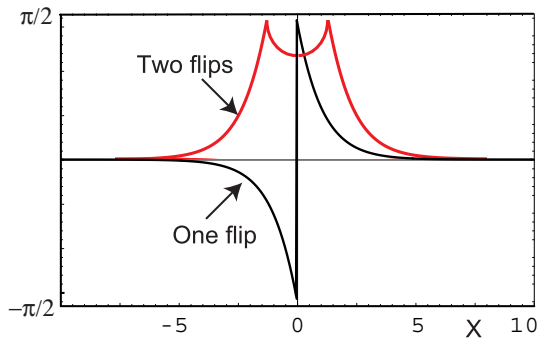}\caption{Phase variation along the
horizontal axis after a flip of a single rhombus (solid curve) and after a
consequitive flips of two rhombi located at a distance twice the core size of
each rhombi, $\sqrt{E_{J}/\epsilon_{J}}$. The horizontal axis shows the
distance, $X=\sqrt{\epsilon_{J}/E_{J}} x$ measured in the units of the vortex
core size.}%
\label{Vortex}%
\end{figure}

We thus begin our analysis of the effects of the finite phase rigidity with
the consideration of the dangerous single rhombus flips. Generally, the
continuous phase can be represented as the sum of two parts: the one that it
is due to the vortices and the spin-wave part which does not change the phase
winding numbers. As usual in XY systems, it is the vortex part that is the
most relevant for the physical properties. In particular, in these arrays it
is the vortex part that controls the dynamics of the discrete subsystem.
Notice that, unlike the conventional arrays, the arrays containing rhombi
allow two types of vortices: half-vortices and full vortices because of the
double periodicity of each rhombi. The flip of the individual rhombi is
equivalent to the creation of the pair of half vortices. If the ground state
of the system contains a liquid of half-vortices, these processes become real
and the main feature of the Hamiltonian, namely the existence of two sets of
anticommuting variables is lost. We now estimate the potential energy of the
half vortex and of the pair associated with single flip, $U_{sf}$, and
amplitude to create such pair, $\widetilde{t}$. We begin with the potential
energy which is different in different limits. Let us consider the simpler
limiting case when rhombi flips do not affect the rigidity in the vertical
direction, it remains $\epsilon_{J}$. Further, we have to distinguish the case
of a very large size in horizontal direction and a moderate size because the
contribution from the individual chains can be domininant in a moderate system
if $E_{J}\gg\epsilon_{J}$. In a very large system of linear size $L$ \ with
rigidity $\epsilon_{J}$ in the vertical direction the potential energy of one
vortex is%
\begin{equation}
E_{v}=\pi\sqrt{E_{J}\epsilon_{J}}\ln(L) \label{E_v}%
\end{equation}
while the energy of the vortex-antivortex pair at a large distance $R$ from
each other is
\begin{equation}
U_{v}(R)=\pi\sqrt{E_{J}\epsilon_{J}}\ln(R) \label{U_v}%
\end{equation}
These formulas can be derived by noting that at large scales the
superconducting phase changes slowly which allows one to use the continuous
approximation for the energy density: $E=\frac{1}{2}E_{J}(\partial_{x}%
\phi)^{2}+\frac{1}{2}\epsilon_{J}(\partial_{z}\phi)^{2}$. Rescaling then the
$x$-coordinate by $x\rightarrow\widetilde{x}\sqrt{E_{J}/\epsilon_{J}}$ we get
an isotropic energy density $E=\frac{1}{2}\sqrt{E_{J}\epsilon_{J}}(\nabla
\phi)^{2}$. The continuous approximation is valid if both rescaled coordinates
$\widetilde{x},z\gtrsim1$. Thus, in a system with $E_{J}\sim\epsilon_{J}$ the
formulas (\ref{E_v},\ref{U_v}) remain approximately correct even at small
distances $R\sim1$, so a flip of a single rhombus creates a halfvortex -
anti-halfvortex pair with energy $E_{p}\approx E_{J}$ but the formulas become
parametrically different in a strongly anisotropic system. Consider first the
limit $\epsilon_{J}=0$. Here the chains of rhombi are completely decoupled and
the energy of two half vortices separated by one rhombus in the vertical
direction (the configuration created by a single flip) is due to the phase
gradients in only one chain, $U_{sf}^{(0)}=\pi^{2}E_{J}/(2L)$, which appear
because the ends of the chain have the phase fixed by the boundary
superconductor. A very small coupling between the chains adds $U_{sf}%
^{(1)}=(2/\pi)\epsilon_{J}L$ to this energy, so the total potential energy of
the single flip inside the array is
\begin{equation}
U_{sf}=\pi^{2}E_{J}/(2L)+(4/\pi)\epsilon_{J}L\qquad L\ll\sqrt{E_{J}%
/\epsilon_{J}} \label{U_sf_L}%
\end{equation}
This formula is correct as long at the second term is much smaller than the
first one; they become comparable at $L=\sqrt{E_{J}/\epsilon_{J}}$ and at
larger $L$ the potential energy associated with the single flip saturates at
\begin{equation}
U_{sf}=\gamma\sqrt{E_{J}\epsilon_{J}}\qquad L\gg\sqrt{E_{J}/\epsilon_{J}}
\label{U_sf}%
\end{equation}
where $\gamma\approx3.3.$ Qualitatively, a single flip leads to the continuous
phase configuration where phase gradients are significant in a narrow strip in
$x$-direction of the length $\sqrt{E_{J}/\epsilon_{J}}$ and width $\sim1$. The
phase configuration resulting from such process is shown in Fig.~\ref{Vortex}.
These formulas assume that the rigidity of the superconducting phase remains
$\epsilon_{J}$ which is, strictly speaking, only true if the discrete
variables are perfectly ordered in the vertical direction. Indeed, the
coupling in the vertical direction contains $\epsilon_{J}\cos(\phi)\tau
_{i,j}^{z}\tau_{i+1,j}^{z}$ which renormalizes to\textbf{\ }$\epsilon_{J}%
\cos(\phi)\langle\tau_{i,j}^{z}\tau_{i+1,j}^{z}\rangle$\textbf{\ }in a
fluctuating system. In the opposite limit of strongly fluctuating rhombi, the
average value of $\langle\tau_{i,j}^{z}\tau_{i+1,j}^{z}\rangle$ becomes small,
we can estimate it from the perturbation theory expansion in $\epsilon_{J}$
which sets the lowest energy scale of the problem: $\langle\tau_{i,j}^{z}%
\tau_{i+1,j}^{z}\rangle\approx\epsilon_{J}/(L_{eff}J_{x})$ which renormalizes
the value of $\epsilon_{J}$:%
\[
\epsilon_{J}\rightarrow\widetilde{\epsilon_{J}}=\frac{\epsilon_{J}^{2}%
}{L_{eff}J_{x}}
\]
This renormalized value of $\epsilon_{J}$ should be used in the estimates of
the vortex energy (\ref{U_sf_L}) or (\ref{U_sf}). This does not affect much
the estimates unless the system is deep in the fluctuating regime.

Unlike potential energy, the single flip processes occur with the amplitude%
\begin{equation}
\widetilde{t}=E_{J}^{3/4}E_{C}^{1/4}e^{-s\sqrt{E_{J}/E_{C}}} \label{tilde_t}%
\end{equation}
in all regimes. This formula neglects the contribution of the continuous phase
to the action of the tunneling process. The reason is that both the potential
energy (\ref{U_sf}) of the half-vortex and the kinetic energy required to
change the continuous phase are much smaller than the corresponding energies
of the individual rhombus, $E_{J},E_{C}$. In order to estimate the kinetic
energy, consider the contribution of the vertical links (horizontal links give
equal contribution). There are roughly $\sqrt{E_{J}/\epsilon_{J}}$ such links,
so their effective charging energy is about $e_{C}\sqrt{\epsilon_{J}/E_{J}}$ .
If all junctions in this array are made with the same technology their
Josephson energies and capacitances are proportional to their areas, so
$\epsilon_{J}/E_{J}=$ $E_{C}/\epsilon_{C}\equiv\eta$, in the following we
shall refer to such junctions as \emph{similar}. In this case the array is
characterized by two dimensionless parameters, $\eta\ll1$ and $E_{J}/E_{C}%
\gg1$ and the additional contribution to the charging energy, $\eta^{1/2}%
E_{C}^{-1},$ coming from vertical links is smaller than the one of the
individual rhombi, $E_{C}^{-1}$ and thus do not change the dynamics.

We conclude that the dangerous real single flip processes become forbidden if
$\widetilde{t}\ll U_{sf}$ where $\widetilde{t}$ is given by (\ref{tilde_t})
and $U_{sf}$ by (\ref{U_sf_L}) or (\ref{U_sf}). This condition is not
difficult to satisfy in a real array because amplitude $\widetilde{t}$ is
typically much smaller than $E_{J}$. Further, for moderately sized arrays
(with $L=5-10$ which already provide a very good protection) the energy of a
single rhombus flip is only slightly smaller than $E_{J}$, so the condition
$\widetilde{t}\ll U_{sf}$ is not really restrictive. Note, however, that in
order to eliminate low energy states of the discrete subsystem we also need to
satisfy the condition (\ref{epsilon_J}) which implies that the tunneling
processes should occur with a significant amplitude. While this might be
difficult in the infinite array made from the similar junctions (with the same
product of charging and Josephson energies), this is not really a restrictive
condition for moderately sized arrays. One can choose, for instance, for a
system of $L\times L$ rhombi with $L=5-10$ Josepshon contacts with
$E_{J}=10E_{C}$. This choice would give $\widetilde{t}\approx0.35E_{C}$ and
$J_{x}\approx0.2E_{C}$ for a system of disconnected horizontal chains. The
condition $\widetilde{t}\ll U_{sf}$ is well satisfied. Choosing now the
vertical links with $\epsilon_{J}=0.5E_{C}$ and corresponding $\epsilon
_{C}=20E_{C}$ we get $L_{eff}\approx5$, so that the condition (\ref{epsilon_J}%
) is satisfied as well and there are no low energy states. It is more
difficult, however, to eliminate the low energy states in the infinite array
of coupled chains shown in Fig. \ref{SArray} and to satisfy the condition
$\widetilde{t}\ll U_{sf}$ at the same time, especially if all junctions are to
be "similar" in the sense defined above. This can be achieved, however, by
replacing the vertical links by rhombi frustrated by the flux $\Phi
=(1/2+\delta)\Phi_{0}$ with $\delta\ll1$ with each junction characterized by
$\widetilde{\epsilon}_{J}\lesssim E_{J}$ and $\epsilon_{C}\gtrsim E_{C}$.
These rhombi would provide a significant rigidity to the continuous phase
fluctuations (with effective rigidity $\widetilde{\epsilon_{J}}$) but only
weak coupling ($\epsilon_{J}=\delta\widetilde{\epsilon_{J}}$) between discrete
degrees of freedom.

Finally, we discuss the effect of the finite phase rigidity on the amplitude
of the two rhombi processes, $J_{x}(k)$. The condition that real single flip
process do not occur does not exclude the virtual processes that flip
consequitively two rhombi in the same chain. This would lead to an additional
contribution to $\widetilde{J_{x}}(k)$ (\ref{J_x(k)}). To estimate this
contribution we note that immediately after two flips the continuous phase has
a configuration shown in Fig.~\ref{Vortex}, which is associated with the
energy $\sim U_{sf}$. The amplitude for such two consequitive flips is
$\widetilde{t}^{2}/U_{sf}$; it can become of the order of $\widetilde{J_{x}%
}(k)$ in a large system (where $U_{sf}$ is small). However, the\ amplitude of
the full process involves additional action which further suppresses this
amplitude. This happens because the two consequitive flips lead to the high
energy virtual state sketched in Fig. \ref{Vortex} and in order to get back to
the low energy state the resulting continuous phase has to evolve dynamically.
To estimate the action corresponding to this evolution, we note that its
dynamics is controlled by $\sqrt{E_{J}/\epsilon_{J}}$ junctions with charging
energy\ $\epsilon_{C}$. For the estimate we can replace these junctions by a
single junction with capacitive energy $\epsilon_{C}\sqrt{\epsilon_{J}/E_{J}}%
$. Thus, the final stage of this process leads to the additional term in the
action $\delta S\sim E_{J}/\epsilon_{C}=\eta(E_{J}/E_{C})$. Depending on the
parameter, $\sqrt{E_{J}E_{C}}/\epsilon_{C}=\eta\sqrt{E_{J}/E_{C}}$, this
additional conribution to the action is smaller or bigger than the total
action, but even if it is smaller, it is still large compared with unity if
$E_{J}\gg\epsilon_{C}$. In this case, the processes that do not change the
continuous phase dominate. We emphasize again that in any case the transitions
involving two flips in the same row commute with both integrals of motion
$P,Q$ and thus do not affect the qualitative conclusions of the previous
section. For practically important similar junctions, it means the following.
If $\eta\gg$ $\sqrt{E_{C}/E_{J}}$ only nearest rhombi flip with the amplitude
$J_{x}$ given by (\ref{J_x(k)}). If $\sqrt{E_{C}/E_{J}}\gg\eta\gg$
$E_{C}/E_{J}$ the flips occur for the rhombi in the same row if they are
closer than $L_{eff}<\eta^{-1}\sqrt{E_{C}/E_{J}}$. Finally for $\eta\ll
E_{C}/E_{J}$ the distance between flipped rhombi exceeds the size of the
half-vortex and the two rhombi flips in a large ($L\gg\sqrt{E_{J}/\epsilon
_{J}}$) array happen via virtual half-vortices in the continuous phase.

In the discussion above we have implicitly assumed a superconducting boundary
conditions such as shown in Fig. \ref{SArray}. These boundary conditions imply
that\ in the absence of significant continuous phase fluctuations $P_{i}%
P_{j}=1$. Physically, it means that if the array as a whole is a
superconductor, it still has two states characterized by the phase difference,
$\Delta\phi=0$ or $\pi$ between the left and the right boundaries even in the
regime where individual phases in the middle fluctuate strongly between values
$0$ and $\pi$. In this regime of strong discrete phase fluctuations, the
external fields are not coupled to the global degree of freedom $\Delta\phi$
describing the array as a whole. In principle, it is also possible to have a
similar array with open boundary conditions but in this case it is more
difficult to eliminate low lying states because there is no reason for the
constraint $P_{i}P_{j}=1$ in this case.

\subsection{Array equivalent to a spin model with local interactions.}

\begin{figure}[th]
\includegraphics[width=3.0in]{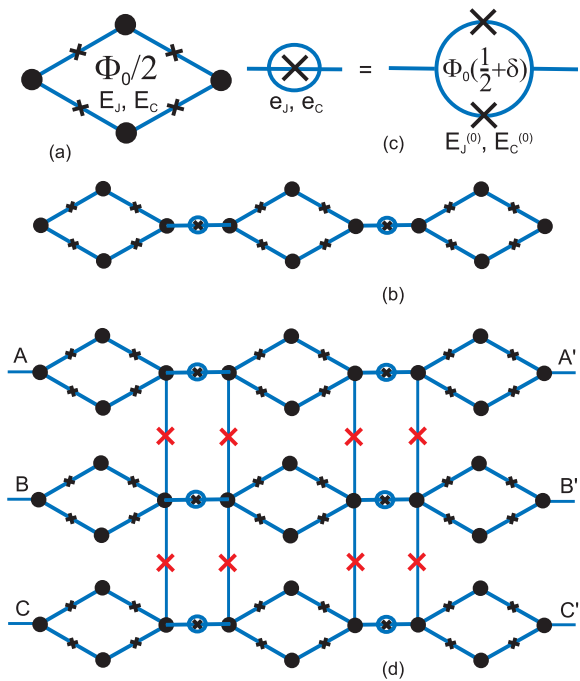}\caption{ Schematics of the array. (a)
The elementary Josephson circuit emulating spin $1/2$ consists of a four
junction loop penetrated by magnetic flux $\Phi_{0}/2$. (b) Implementation of
the spin chain by Josephson junction loops, here elementary rhombi are
connected by weak links, the appropriate parameters for these links can be
obtained if each link in fact consists of two elementary Josephson junctions
as shown in (c). (d). Full array implementing the spin model (\ref{H}). The
natural boundary conditions for this array are periodic, i.e. the point $A$
should be connected to $A^{\prime}$, $B$ to $B^{\prime}$, etc. }%
\label{FullArray}%
\end{figure}

In order to construct the array equivalent to the spin model (\ref{H}) we need
to couple the rhombi in such a way that the transitions involving only one
rhombus in a row are not allowed but the superconducting phase varies
significantly between one rhombus and the next in the row. This is achieved if
rhombi are connected in a chain by a weak Josephson link, characterized by
Josephson energy $e_{J}$ and Coulomb energy $e_{C}$, so that $E_{J}\gtrsim
e_{J}$ and $e_{C}\ll E_{C}$, as shown in Fig.~\ref{FullArray}b. In this case
the simultaneous tunneling of two rhombi which does not change the phase
difference across the weak junction is not affected, its amplitude, $J_{x}$,
is still given by (\ref{J_x(k)}).

In this array the flip of a single rhombus is due to two alternative
mechanisms. The first one is that it involves the creation of the half-vortex
(as discussed in the previous subsection)\ and does not involve the change in
the phase accross the weak junction. Alternatively, it can be due to the phase
flip by $\pi$ across the weak junction. Because $e_{C}\ll E_{C}$ this process
is slow and its amplitude is low:
\[
t=E_{J}^{3/4}e_{C}^{1/4}e^{-s^{\prime}\sqrt{E_{J}/e_{C}}}
\]
where $s^{\prime}\approx1$. Further, this process increases the energy of the
system by $e_{J}$, so if its amplitude is small: $t\ll e_{J}$ it can be
completely neglected. Normally junctions with smaller Josephson energy,
$e_{J}\lesssim E_{J}$, have a bigger charging energy, not a smaller one as
required here. To avoid this problem, we note that these weak junctions can be
in practice implemented as a two junction SQUID loops containing the flux
$\Phi=(1/2+\delta)\Phi_{0}$ as shown in Fig.~\ref{FullArray}c. The effective
Josephson coupling provided by such loop is $e_{J}=\sin(2\pi\delta)E_{J}%
^{(0)}$ (where $E_{J}^{(0)}$ is the energy of the individual junction) while
its effective charging energy is $e_{C}=E_{J}^{(0)}/2$. This allows to use
bigger junctions for these weak junctions and has another advantage that it
provides an additional controlling parameter on the system. This construction
seems somewhat similar to the partially frustrated rhombi that one needs to
introduce as vertical links in large arrays discussed in previous subsection
(Fig. \ref{SArray}c) but it serves a completely opposite purpose: it increases
the density of the full vortices while keeping the discrete variables coupled.
The partially frustrated rhombi, on the other hand, suppress the fluctuations
of the continous phase while allowing independent fluctuations of the discrete
variables in different rows. Such a dramatic difference is made possible by a
combination of two reasons. First, the frustration has a different effect on
these elements: in the case of the SQUID loops half flux eliminates the
Josephson coupling completely while in a case of rhombi it leads to the exact
degeneracy between two discrete states and to a rigidity for the continuous
phase. Second, the values of the charging energies are rather different: in
the case of SQUID loops the charging energy of its junctions completely
dominates all processes in which phase changes across this loop thereby
prohibiting the single rhombus flips. In the case of partially frustrated
rhombi their charging energy is of the order of the charging energy of the
contacts in horizontal rhombi and thus it suppresses the double flips of
distant rhombi in one chain but has relatively minor effect on the nearest ones.

The $J_{z}$ interaction between the spins is provided by the pairs of the weak
Josephson contacts as shown in Fig.~\ref{FullArray}d with the Josephson energy
$\epsilon_{J}\lesssim E_{J}$ and charging energy $\epsilon_{C}\gtrsim E_{C}$.
These junctions do not affect the tunneling process of each rhombi but provide
the weak interaction between them of the form (\ref{H}) with the strength
\begin{equation}
J_{z}=\epsilon_{J} \label{J_z}%
\end{equation}
Similar to the array discussed in the previous subsection we have to choose
the parameters so that the energy of the half vortex and states resulting from
a single flip is sufficiently high compared to the amplitude of the single
flips. The discussion of the previous subsection carries over to this array.
The only change is that the energy of the vortex in the infinite system
contains the weakest link in the horizontal direction, i.e. $e_{J}$ instead of
$E_{J}$:
\begin{align*}
\widetilde{t}  &  \ll U_{sf}\\
&
\begin{tabular}
[c]{ll}%
$U_{sf}=\pi^{2}e_{J}/(2L)+(4/\pi)\epsilon_{J}L$ & $L\ll\sqrt{e_{J}%
/\epsilon_{J}}$\\
$U_{sf}=\gamma\sqrt{e_{J}\epsilon_{J}}$ & $L\gg\sqrt{e_{J}/\epsilon_{J}}$%
\end{tabular}
\qquad
\end{align*}
Finally, we have to ensure that the phases of the consequitive rhombi are
decoupled and the interaction between discrete degrees of freedom is purely
local. This is satisfied if the continuous phase across weak junctions
fluctuates strongly, i.e. that the energy of a usual (not half)\ vortex is
smaller than the kinetic energy: $\sqrt{e_{J}\epsilon_{J}}\ll e_{C}$.
Physically, it means that the array as a whole is an insulator, similar to the
topological insulator considered in \cite{Doucot2003} due to the full vortices
that move in a vertical direction thereby decoupling different columns of
rhombi. This condition does not contradict the condition $\widetilde{t}\ll
U_{sf}$ because the latter involves the exponentially small amplitude of
flipping a single rhombus. If both conditions are satisfied the absolute value
of the phase on each island constituting a rhombus fluctuates but the
difference across the rhombus remains a slow varible taking two discrete
values and it flips only simultaneously with the another one in the same row.
Note that the interaction between discrete variables belonging to one column
is due to the loop formed by these rhombi and two vertical junctions, it is
therefore always local by construction, its value is given by (\ref{J_z}).
Repeating the arguments of the previous subsection, we see that in order to
suppress the simultaneous flips of distant rhombi in the same row one needs
also to satisfy the condition $E_{J}E_{C}/\epsilon_{C}^{2}\gg1$, but in
contrast to the case of the regime discussed there, here the conditions on the
vertical junctions are not difficult to satisfy because one does not need to
keep the long range order in a continuous phase. Under these conditions the
dynamics of the array is described by the Hamiltonian (\ref{H}).

Although in this regime the system as a whole is an insulator, it does not
allow a half vortex to move across. So, physically, the two states of the
global system can be observed in the array with the periodic boundary
conditions shown in Fig \ref{FullArray}: here two different states correspond
to the half vortex trapped or not trapped inside the big loop formed by the
array as a whole due to the periodic boundary conditions.

\section{Lattice Chern-Simons gauge theories with a finite Abelian group}

\label{CS}

In this section we discuss the general properties of the Chern-Simons theories
with discrete gauge group ${\mathchoice {\hbox{$\sf\textstyle
Z\kern-0.4em Z$}} {\hbox{$\sf\textstyle Z\kern-0.4em Z$}} {\hbox{$\sf\scriptstyle
Z\kern-0.3em Z$}} {\hbox{$\sf\scriptscriptstyle Z\kern-0.2em Z$}}}_{n}$.
\emph{Continuous} Chern-Simons theories recieved a lot of attention in the
past because they provide a natural mechanism for the flux attraction to the
charged particles which thereby change their statistics \cite{Arovas85}. In
spite of some technical difficulties, it is possible to construct lattice
versions of Chern-Simons theories \cite{Eliezer92,Diamantini93}, at least for
\emph{continuous} groups. The case of a \emph{discrete} gauge group is
slightly more delicate bcause one can not define canonically conjugate pairs
such as the ``magnetic'' ($A_{ij}$) and the ``electric'' ($\frac{\partial
}{i\partial A_{ij}}$) variables. Before we discuss the peculiarities
associated with discrete groups, we briefly recapitulate the main properties
of the continuous Chern-Simons theories with the non-compact and compact
$U(1)$ groups which we describe in the Hamiltonian formalism that we need to
generalize these models to discrete groups. The $U(1)$ Chern-Simons model is
usually described by the Lagrangian:
%
\begin{equation}
\mathcal{L} = \frac{1}{2}\lambda(\dot{A}_{x}^{2} + \dot{A}_{y}^{2}) - \frac
{1}{2}\mu B^{2} + \nu(\dot{A}_{x}A_{y}-\dot{A}_{y}A_{x}) \label{L_CS}%
\end{equation}
where $B=\partial_{x}A_{y}-\partial_{y}A_{x}$ and dots stand for
time-derivatives. We have used the gauge in which the time component $A_{0}$
of the vector potential is zero. Because of this, we shall use only invariance
under time-independent gauge transformations in this discussion. The canonical
variables conjugate to $A_{x}$ and $A_{y}$ are respectively $\Pi_{x}=
\lambda\dot{A}_{x} + \nu A_{y}$ and $\Pi_{y} =\lambda\dot{A}_{y} - \nu A_{x}$.
The gauge transformations of the classical fields $A_{\rho}$ are the usual
ones $A_{\rho}\rightarrow A_{\rho}+ \partial_{\rho}\phi$, but because of the
Chern-Simons term this also induces a transformation of the conjugate fields
$\Pi_{\rho}\rightarrow\Pi_{\rho}+ \nu\epsilon_{\rho\sigma} \partial_{\sigma}
\phi$, where $\epsilon_{xy}=-\epsilon_{yx}=1$.

In quantum theory, $\Pi_{\rho}$ and $A_{\rho}$ become operators with the
commutation relation $[\Pi_{\rho},A_{\mu}]=-i\delta_{\rho,\mu}(r-r^{\prime})$
and the gauge transformation is generated by the operator $R_{\phi}$ defined
by $R_{\phi}=\int d^{2}rR(r)\phi(r)$ with $R(r)=\partial_{\rho} \Pi_{\rho}(r)
+ \nu B(r)$, since $[R(r),A_{\rho}(r^{\prime})]=-i\partial_{\rho}%
\delta(r-r^{\prime})$ and $[R(r), \Pi_{\rho}(r^{\prime})]= -i \nu
\epsilon_{\rho\sigma} \partial_{\sigma}\delta(r-r^{\prime})$. In more physical
terms, introducing an ``electric'' field (which is equal to $\lambda\dot{A}$)
in the classical theory) by $E_{x} = \Pi_{x}-\nu A_{y}$ and $E_{y} = \Pi
_{y}+\nu A_{x}$, the generator of the gauge transformations can be expressed
by $R(r)=\partial_{\rho}E_{\rho}+ 2 \nu B$. It is simple to check that the
gauge transformations commute among themselves and also commute with the
Hamiltonian density $\mathcal{H}=\frac{1}{2\lambda}E^{2}-\frac{1}{2}\mu B^{2}%
$. In the absence of matter, the equations of motion for the fields read:
%
\begin{equation}
\dot{E}_{\rho}+\epsilon_{\rho\sigma}(\frac{2\nu}{\lambda}E_{\sigma}
+\mu\partial_{\sigma}B) = 0
\end{equation}
\begin{equation}
\lambda\dot{B} = \epsilon_{\rho\sigma} \partial_{\rho}E_{\sigma}%
\end{equation}
This yields a massive branch of excitations, which are no longer purely
transverse, but also develop a longitudinal component proportional to $\nu$.
Their dispersion relation is $\lambda^{2}\omega^{2}_{k}=4\nu^{2}+\lambda\mu
k^{2}$. These propagating modes are then pushed to very high energies in the
limit where $\lambda\ll\nu$. In the presence of static external charges
$e_{i}$ at positions $r_{i}$, the Gauss constraint becomes: $R(r)=\sum_{i}
e_{i}\delta(r-r_{i})$. In the ground state, these charges induce a static
field configuration according to: $(2\nu-\frac{\lambda\mu}{2\nu}\nabla
^{2})B(r)=\sum_{i} e_{i}\delta(r-r_{i})$. So each particle is bound to a flux
tube carrying a flux equal to $\frac{e_{i}}{2\nu}$ and smeared over a typical
length $\xi=\frac{(\lambda\mu)^{1/2}}{\nu}$. The pure Chern-Simons limit is
then recovered as $\xi$ goes to zero. In this limit, when one particle of
charge $e_{1}$ goes around an other charge $e_{2}$, the total wave-function of
the system is multiplied by an Aharonov-Bohm phase factor equal to
$\exp(i\frac{e_{1}e_{2}}{2\nu})$, so we a get a factor $\exp(i\frac{e_{1}%
e_{2}}{4\nu})$ upon exchanging these two particles. We note that in the limit
$\mu\rightarrow0$ $H$ commutes with operators $\tilde{\Pi}_{\mu} ={\Pi}_{\mu}
+ \nu\epsilon_{\mu\eta} A_{\eta}$. These operators provide a generalization of
the usual shift operators ($\Pi_{\mu}$) for the Chern-Simons theory.
Physically, the additional term in these operators appears because in the
presence of Chern-Simons one can not only shift fields, one has also to change
the phase of the wave function accordingly.

\begin{figure}[th]
\includegraphics[width=3.0in]{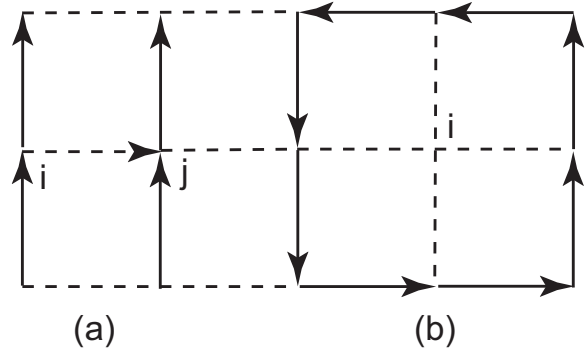}\caption{ (a) An oriented link $(ij)$
(dashed line) and the four oriented links adjacent to it (thick lines) which
enter in the set $\mathcal{N}(ij)$. (b) For a given site $i$, the loop
$\mathcal{L}(i)$ defined in text (thick lines). }%
\label{Definitions}%
\end{figure}

We now turn to the lattice versions of the Chern-Simons theory. For the sake
of simplicity, we shall work here on a square lattice, although these
constructions could be generalized to other periodic systems. In this case,
the vector potential describing the gauge field is defined on the links of the
lattice, and will be denoted by $A_{ij}$ for the oriented link connecting
sites $i$ and $j$. If we reverse the orientation of the link, we obtain
\mbox{$A_{ji}=-A_{ij}$}. We shall adopt the Hamiltonian description from now
on. In the absence of a Chern-Simons term, the local electric fields are
simply the canonical conjugate operators $\Pi_{ij}$ of the $A_{ij} $'s. The
corresponding commutation relations become:
%
\begin{align*}
\;[A_{ij},A_{kl}]  &  = 0\\
\;[\Pi_{ij},\Pi_{kl}]  &  = 0\\
\;[A_{ij},\Pi_{kl}]  &  = i\delta_{(ij),(kl)}%
\end{align*}
The natural lattice Hamiltonian whose continuum limit is the same as before
reads:
%
\begin{equation}
H= \frac{1}{2\lambda}\sum_{<ij>}\Pi_{ij}^{2} +\frac{\mu}{2} \sum_{ijkl}%
(A_{ij}+A_{jk}+A_{kl}+A_{li})^{2} \label{H0noncompact}%
\end{equation}
In the presence of a Chern-Simons term, the electric field operators are
modified. The electric field $E_{ij}$ will contain, besides $\Pi_{ij}$, terms
associated to the vector potential in the direction perpendicular to the link
$(ij)$. On a square lattice, there are four links immediately perpendicular to
this link, and containing either site $i$ or $j$. Let us denote by
$\mathcal{N}(ij)$ this set of four links. To reflect the signs which appear in
the continuous version discussed above, the links in $\mathcal{N}(ij)$ have to
be oriented in such a way as when $(ij)$ runs from left to right, $(kl)$ runs
from bottom to top, as illustrated in Fig.~\ref{Definitions}(a) below.

With these notations, the lattice Hamiltonian in the presence of a
Chern-Simons term may be written as:
%
\begin{align}
H_{CS}  &  = \frac{1}{2\lambda}\sum_{<ij>}\left(  \Pi_{ij}-\frac{\nu}{4}
\sum_{(kl)\in\mathcal{N}(ij)}A_{kl}\right)  ^{2}\nonumber\\
&  \mbox{} +\frac{\mu}{2} \sum_{ijkl}(A_{ij}+A_{jk}+A_{kl}+A_{li})^{2}
\label{HCSnoncompact}%
\end{align}
The important fact is that the generalized electrical field operators
\mbox{$E_{ij}=\Pi_{ij}-\frac{\nu}{4}\sum_{(kl)\in\mathcal{N}(ij)}A_{kl}$} no
longer commute. The relation $[E_{x}(r),E_{y}(r^{\prime})]=-2i\nu
\delta(r-r^{\prime})$ from the continuous case becomes now: $[E_{ij}%
,E_{kl}(r^{\prime})]=-\frac{i}{2}\nu$, whenever $(kl)$ is one of the four
links in $\mathcal{N}(ij)$. Here $(ij)$ is oriented along the positive
$\mathbf{x}$ axis, and $(kl)$ along the positive $\mathbf{y}$ axis. The
apparent difference in the normalization of the delta function on the right
hand sides of these expressions is compensated by the fact that $\mathcal{N}%
(ij)$ contains four elements.

The energy spectrum for the eigenmodes associated to this system now reads:
\[
\lambda^{2}\omega_{k}^{2}=4\nu^{2}\cos^{2}(\frac{k_{x}}{2})\cos^{2}%
(\frac{k_{y}}{2})+4\lambda\mu(\sin^{2}(\frac{k_{x}}{2})+\sin^{2}(\frac{k_{y}%
}{2}))
\]
This spectrum is specially interesting in the limit where $\mu$ goes to zero,
since then it exhibits lines of zero modes at the boundaries of the first
Brillouin zone, namely for $k_{x}=\pm\pi$ or $k_{y}=\pm\pi$. These modes are
directly related to two remarkable non-local conservation laws which appear in
the limit of vanishing $\mu$. More precisely, for each row and each column, we
may define two conserved quantities $\tilde{P}_{\mathrm{row}}$ and $\tilde
{Q}_{\mathrm{column}}$ in the following way:
%
\begin{align}
\tilde{P}_{\mathrm{row}}  &  =\sum_{j\in\mathrm{row}}(-1)^{x_{j}%
}E_{j,j+\mathbf{\hat{x}}}\label{constcontP}\\
\tilde{Q}_{\mathrm{column}}  &  =\sum_{j\in\mathrm{column}}(-1)^{y_{j}%
}E_{j,j+\mathbf{\hat{y}}} \label{constcontQ}%
\end{align}
%
Unlike the row and column operators discussed in Section II these operators
mutually commute. Note that in the \mbox{$\mu =0$} limit, the system exhibits
a large set of \emph{local} conserved quantities $\tilde{\Pi}_{ij}$, i.e. one
for each link, defined by:
\begin{equation}
\mbox{$\tilde{\Pi}_{ij}=\Pi_{ij}+\frac{\nu}{4}\sum_{(kl)\in\mathcal{N}(ij)}
A_{kl}$}. \label{tilde_Pi}%
\end{equation}
Similar to the local electric fields, these variables do not mutually commute,
so they cannot be simultaneously diagonalized. Furthermore, they are not
invariant under gauge transformations, since the electric field opeators are
gauge invariant and since
\mbox{$E_{ij}-\tilde{\Pi}_{ij}=-\frac{\nu}{2}\sum_{(kl)\in\mathcal{N}(ij)}
A_{kl}$} which is clearly gauge dependent. This situation is very similar to
what occurs in the process of quantizing the motion of a particle on a plane
in the presence of a uniform perpendicular magnetic field. The local
electrical field operators are analogous to the gauge invariant velocity
operators, $P-eA$, in the particle problem. Because of the magnetic field, the
two components of this vector no longer commute. Since the vector potential
$A$ is not translation invariant, the usual translation operators have to be
combined with gauge transformations in order to commute with the kinetic
energy $(P-eA)^{2}$. These deformed generators of translations are analogous
to the two components of the vector $\tilde{\Pi}=P+eA$. Apparently, the only
way to construct gauge-invariant symmetry operators in the lattice gauge model
is to use the non-local combinations $\tilde{P}_{\mathrm{row}}$ and $\tilde
{Q}_{\mathrm{column}}$ defined in (\ref{constcontP},\ref{constcontQ}). Note
that a model with similar conservation laws, for interacting Bosons on a
square lattice, has been analyzed in~\cite{Paramekanti02}.

A next step is to construct a lattice Chern-Simons gauge theory for the
continuous but \emph{compact} $U(1)$ group. This is simply achieved in the
absence of a Chern-Simons term, by assuming that the vector potential
variables $A_{ij}$ are \emph{periodic}, with a period chosen for instance
equal to $2\pi$. This implies that the spectrum of the conjugate operators
$\Pi_{ij}$ are \emph{discrete}, containing only integer values. Because of
this, the form of the Hamiltonian has to be modified from
equation~(\ref{H0noncompact}) above, and a natural choice respecting the
requirements of gauge invariance and periodicity in the gauge potentials
reads:
%
\begin{equation}
H=\frac{1}{2\lambda}\sum_{<ij>}\Pi_{ij}^{2}-\mu\sum_{ijkl}\cos(A_{ij}%
+A_{jk}+A_{kl}+A_{li})^{2} \label{H0compact}%
\end{equation}
%
Using this Hamiltonian as a starting point we add a Chern-Simons term by the
standard deformation of the electrical operators:
%
\begin{align}
H_{CS}  &  =\frac{1}{2\lambda}\sum_{<ij>}\left(  \Pi_{ij}-\frac{\nu}{4}%
\sum_{(kl)\in\mathcal{N}(ij)}A_{kl}\right)  ^{2}\nonumber\\
&  \mbox{}-\mu\sum_{ijkl}\cos(A_{ij}+A_{jk}+A_{kl}+A_{li})^{2}
\label{HCScompact}%
\end{align}
%
We now discuss what can be the Hilbert space associated with this Hamiltonian.
This is a non-trivial problem because we can no longer impose the periodicity
in the local gauge potentials in the usual way, assuming that the
wave-function of the system considered as a function of the $A_{ij}$'s is
$2\pi$ periodic with respect to any $A_{ij}$. This is not possible because the
naive shift operator $S_{ij}^{naive}=\exp(-i2\pi\Pi_{ij})$ that changes
$A_{ij}$ into $A_{ij}+2\pi$ no longer commutes with the kinetic part of the
Hamiltonian. The appropriate definition of these shift operators becomes
$S_{ij}=\exp(-i2\pi\tilde{\Pi}_{ij})$, with the $\tilde{\Pi}_{ij}$ defined in
(\ref{tilde_Pi}). It order to construct invariant states under this full set
of shift operators, we need them to be mutually commuting. This is realized
only for a discrete set of values of $\nu=\frac{m}{\pi}$, where $m$ is any
integer. Thus, the compact gauge theory is compatible with the Chern Simons
term only for special, \textquotedblright quantized\textquotedblright, values
of $\nu$ ~(see also \cite{Alvarez85,Henneaux86}) This statement is very
similar to the well-known fact that in order to quantize the problem of a
particle on a torus in a perpendicular uniform magnetic field, the total
magnetic flux through this torus should be an integer mutiple of the flux
quantum $\frac{h}{e}$. In one particle problem this requirement simply
expresses the need for mutual commutation between two magnetic translations
that are used to contruct the torus from an infinite plane.

Note that, even for these special values of $\nu$, the Hilbert space of the
theory is peculiar because the shift operators $S_{ij}$ are not gauge
invariant. More precisely, let us denote the generator of the gauge
tranformation sending $A_{ij}$ into $A_{ij}+\alpha$, for any site $j$
connected to site $i$ by \mbox{$U_{i}(\alpha)=\exp(-i\alpha
\sum_{j}^{(i)}\tilde{\Pi}_{ij})$}. In this expression, the sum is taken over
the nearest neighbors $j$ of site $i$. Clearly, these unitary operators
commute with the Hamiltonian~(\ref{HCScompact}). For a site $i$, and a link
$(jk)$ belonging to $\mathcal{L}_{i}$, where $\mathcal{L}_{i}$ is the oriented
loop (see Fig.~\ref{Definitions}(b)) defined by the edges of the square built
from four elementary plaquettes whose center is located at site $i$, we have
the following relations:
%
\begin{equation}
U_{i}(\alpha)S_{jk}=e^{\pm i\pi\nu\alpha}S_{jk}U_{i}(\alpha)
\label{commshiftgauge}%
\end{equation}
%
The sign in the phase factor depends on the orientation of the link $(jk)$,
and is negative if it is oriented along $\mathcal{L}_{i}$. This equation
implies that \emph{one cannot enforce at the same time a condition of gauge
invariance on the physical states, and invariance of the wave-function under
the }$2\pi$\emph{\ shift operators}. If we choose to work in a basis of
eigenvectors for the gauge transformations, namely states $|\Psi\rangle$
satisfying \mbox{$U_{i}(\alpha)=\exp(i\alpha Q_{i})|\Psi\rangle$}, where
$Q_{i}$ is the electrical charge at site $i$, applying $S_{jk}$ to
$|\Psi\rangle$ for $(jk)$ in $\mathcal{L}_{i}$ modifies the charge $Q_{i}$ by
$\pm m$ if $\nu=\frac{m}{\pi}$. Conversely, a subspace where $S_{jk}%
|\Psi\rangle=e^{i\theta_{jk}}|\Psi\rangle$ is \emph{not} gauge-invariant,
since after applying $U_{i}(\alpha)$ on such states, we get the new periodic
conditions with \mbox{$\theta'_{jk}=\theta_{jk}+\pi\nu\alpha$}, for any link
$(jk)$ belonging to $\mathcal{L}_{i}$ with the corresponding orientation.
However, it appears that $\sum_{k}^{(j)}\theta_{jk}^{\prime}=\sum_{k}%
^{(j)}\theta_{jk}$ because $\prod_{k}^{(j)}S_{jk}=U_{j}(2\pi)$ commutes with
any gauge transformation. This relation shows that the gauge invariant
quantity $\sum_{k}^{(j)}\theta_{jk}$ is nothing but $2\pi Q_{j}$ modulo $2\pi$.

The properties of this model have been investigated by several groups. In the
absence of a Chern-Simons term, the periodicity of the $U(1)$ gauge field
allows quantum tunneling processes where the local flux on a plaquette changes
by $\pm2\pi$. In a $2+1$ dimensional path integral description, these
instantons (called here monopoles) interact via a Coulomb-like $1/r$
potential, leading to Debye screening in this monopole plasma. The
proliferation of such tunneling events drives the system into a phase where
the magnetic variables $A_{ij}$ are strongly disordered, and in which external
static electric charges interact by a confining potential increasing linearly
with their separation~\cite{Polyakov75,Polyakov77}. In the presence of a
Chern-Simons term, the properties of the monopole plasma are deeply altered,
and several groups have reached the conclusion that a linear interaction now
binds pairs of monopoles of opposite charges, thus destroying the confinement
of electrical charges~\cite{Pisarski86, Affleck89,Diamantini93}.

To extend this construction to a discrete ${\mathchoice{\hbox{$\sf\textstyle
Z\kern-0.4em Z$}}{\hbox{$\sf\textstyle Z\kern-0.4em
Z$}}{\hbox{$\sf\scriptstyle
Z\kern-0.3em Z$}}{\hbox{$\sf\scriptscriptstyle Z\kern-0.2em Z$}}}_{n}$ group,
we replace the continuous vector potential on the link joining sites $i$ and
$j$ by $A_{ij}=\frac{2\pi}{n}p_{ij}$, where $p_{ij}$ is an integer. In the
absence of a Chern-Simons term, the generator of the gauge transformation
based at site $i$ sending $p_{jk}$ into $p_{jk}+\delta_{ji}-\delta_{ki}$ is
$U_{i}=\prod_{j}^{(i)}\pi_{ij}^{+}$, in analogy with the continuous case
discussed above. The unitary operator $\pi_{ij}^{+}$ is analogous to the
operator $\exp(-i\frac{2\pi}{n}\Pi_{ij})$ of the continuous model, namely it
transforms $A_{ij}$ into $A_{ij}+2\pi/n$ or equivalently $p_{ij}$ into
$p_{ij}+1$. In order to attach flux to particles, the generator $U_{i}$ has to
be modified by a phase factor $\exp(-i\frac{\nu}{4}(\frac{2\pi}{n})^{2}%
\sum_{(jk)\in\mathcal{L}_{i}}p_{jk})$, where the $(jk)$'s belong to the loop
$\mathcal{L}_{i}$ already defined above (see Fig.~\ref{Definitions}(b)). For
each value of $\nu$, we define the generators by:
%
\begin{equation}
U_{i}=\prod_{j}^{(i)}\pi_{ij}^{+}\exp(-i\frac{\nu}{4}(\frac{2\pi}{n})^{2}%
\sum_{(jk)\in\mathcal{L}(i)}p_{jk}) \label{DefUi}%
\end{equation}
%
This definition preserves the fact that these generators mutually commute.
Note that since the generators $U_{i}$ commute simultaneously with the local
fluxes, a convenient gauge-invariant basis of the Hilbert space is obtained by
simultaneous diagonalization of the local fluxes and the $U_{i}$'s. With a
discrete basis to describe each link, the kinetic part in the
Hamiltonian~(\ref{HCScompact}) has to be modified. The most natural way to do
this is to replace the local electrical field $E_{ij}$ by a gauge-invariant
operator $\mathcal{E}_{ij}^{+}$ which shifts $A_{ij}$ by the minimal possible
amount $2\pi/n$. This operator is defined as follows:
\[
\mathcal{E}_{ij}^{+}=\pi_{ij}^{+}\exp(i\frac{\nu}{4}(\frac{2\pi}{n})^{2}%
\sum_{(kl)\in\mathcal{N}(ij)}p_{kl})
\]
The Hamiltonian may now be written as:
%
\begin{align}
H_{CS}  &  =-\frac{1}{{\lambda_{n}}}\sum_{<ij>}(\mathcal{E}_{ij}%
^{+}+\mathcal{E}_{ij}^{-})\nonumber\\
&  \mbox{}-\mu\sum_{ijkl}\cos(\frac{2\pi}{n}(p_{ij}+p_{jk}+p_{kl}+p_{li}))^{2}
\label{HCSZn}%
\end{align}
%
where $\mathcal{E}_{ij}^{-}$ is the Hermitian conjugate of $\mathcal{E}%
_{ij}^{+}$. To recover the Hamiltonian~(\ref{HCScompact}) in the limit where
$n$ becomes very large, we notice that $\mathcal{E}_{ij}^{+}$ acts in the same
way as $\exp(-i\frac{2\pi}{n}E_{ij})$, therefore we have to choose
$\lambda_{n}$ so that $2(2\pi/n)^{2}\lambda_{n}^{-1}=\lambda^{-1}$.

With these notations, the operators shifting the link variables $p_{ij}$ by
one unit, and which commute with all the gauge-invariant kinetic terms
$\mathcal{E}_{ij}^{\pm}$ read:
\[
\tilde{\pi}_{ij}^{+}=\pi_{ij}^{+}\exp(-i\frac{\nu}{4}(\frac{2\pi}{n})^{2}%
\sum_{(kl)\in\mathcal{N}(ij)}p_{kl})
\]
since they are analogous to $\exp(-i\frac{2\pi}{n}\tilde{\Pi}_{ij})$ in the
continuous model. The $2\pi$ shift operators $S_{ij}$ previously introduced
are then equal to $(\tilde{\pi}_{ij}^{+})^{n}$. Note that the parameter $\nu$
is quantized, in the same way as before (namely $\nu=\frac{m}{\pi}$, with
integer $m$), since we impose the model to be periodic when $p_{ij}$ is
changed into $p_{ij}+n$. More precisely, as for the compact $U(1)$ group, this
notion of periodicity requires the mutual commutation between all the $2\pi$
shift operators $S_{ij}$. Models obtained from two values of $\nu$ which
differ by an integer multiple of $2n^{2}/\pi$ are clearly identical. We also
note that changing $m$ into $2n^{2}-m$ amounts to replacing all the phase
factors entering in the definition of various operators such as $U_{i}$ by
their complex conjugates, and the corresponding models exhibit similar
properties. In the special case $m=n^{2}$, the operators $\mathcal{E}_{ij}%
^{+}$ mutually commute, so this case is equivalent to $m=0$. Therefore, it is
sufficient to choose $m$ in the set of integers between 0 and $n^{2}-1$. Among
those $n^{2}$ possible values of $\nu$, there is an interesting subset of $n$
values for which $m$ is an integer multiple of $n$. If this condition holds,
the generators of the elementary gauge transformations $U_{i}$ defined in
equation~(\ref{DefUi}) commute with all the $2\pi$ shift operators $S_{jk}$.
It is then possible to apply \emph{simultaneously} the condition of gauge
invariance and the periodic boundary conditions $S_{jk}|\Psi\rangle
=|\Psi\rangle$ for the allowed physical states.

If $\nu=0$, the resulting discrete gauge theory (without Chern Simons term)
has two regimes according to the relative size of the two terms entering in
$H_{CS}$. When $\mu\gg\lambda_{n}^{-1}$, $H(\nu=0)$ describes a phase where
localized flux excitations, called fluxons, have the energy gap $\mu
(1-\cos(\frac{2\pi}{n}))$. In this phase, quantum fluctuations of the magnetic
variables $p_{ij}$ are small, and the effective interaction between two static
external electric charges varies logarithmically with their separation. Notice
that this phase owes its existence to the \emph{discrete} nature of the
symmetry group, as illustrated by the vanishing of the corresponding energy
gap as $n$ is taken to infinity . A small $\lambda_{n}^{-1}$ term simply gives
some amount of dispersion to these excitations. When $\lambda$ is decreased
further below a critical value $\lambda_{n}^{c} $, the fluxon gap closes, and
the system enters the charge confining disordered phase similar to the one
obtained in the \emph{compact} $U(1)$ case, for all values of $\lambda$.

Let us first consider the effect of a switching on a Chern-Simons term in the
former regime, where the potential energy (proportional to $\mu$) dominates.
When $\lambda_{n}$ is very large, we do not expect the flux binding mechanism
to operate. Indeed, a unit charge at site $j$ corresponds to imposing the
Gauss law constraint $U_{k}|\Psi\rangle=\exp(i\frac{2\pi}{n}\delta_{jk}%
)|\Psi\rangle$. When $\lambda_{n}^{-1}$ is small, it is energetically more
favorable to keep a vanishing flux everywhere, because of the low value of the
kinetic term compared to the fluxon gap. So we expect the flux attachment
mechanism to work only if $\lambda_{n}$ is smaller than a critical value
$\lambda_{n}^{\ast}$. When $\lambda_{n}$ is further reduced, below
$\lambda_{n}^{c}$, the fluxon gap eventually closes, and the qualitative
properties of the system change drastically. In Section V we present a simple
perturbative estimate of the critical value $\lambda_{n}^{c}$ and argue that
it is in fact equal to $\lambda_{n}^{\ast}$, i.e. both transitions happen
simultaneously. As already discussed for the case of the compact $U(1)$ group,
the presence of the Chern-Simons term modifies deeply the picture obtained for
vanishing $\nu$ in the strongly fluctuating regime of small $\lambda_{n}$.
Note that, by contrast to the $\nu=0$ case, analysis of the $\lambda
\rightarrow0$ limit is difficult since the $E_{ij}$ operators no longer
commute if they involve two perpendicular links sharing a common site. In the
$n\rightarrow\infty$ limit, we expect to recover the continuous, but
\emph{compact} $U(1)$ theory for which we still do not know how to write down
explicitely the ground state wave-function, even in the $\lambda\rightarrow0$ limit.

\section{Mapping Chern-Simons theories onto spin models}

Here we shall study in more detail the interesting limit of the vanishing
magnetic energy and show explicitely how, in the $n=2$ case, it maps precisely
on the models studied in the beginning of this paper. As we have discussed, it
is possible to propose a design of Josephson-Junction arrays which directly
implements this limit.

As a first step, it is convenient to introduce a basis in the Hilbert space of
gauge invariant states (i.e. states $|\Psi\rangle$ such that $U_{i}%
|\Psi\rangle=0$ for any site $i$), which keeps track of the flux variables.
For any square plaquette $(ijkl)$ centered at $\mathbf{r}$, the corresponding
flux $\sigma_{\mathbf{r}}$ is the integer
\mbox{$p_{ij}+p_{jk}+p_{kl}+p_{li}$}. For any flux configuration
$\{\sigma_{\mathbf{r}}\}$, we may then define a gauge-invariant quantum state
$|\Psi(\{\sigma_{\mathbf{r}}\})\rangle$. Our main task is to represent the
algebra of gauge invariant operators $\mathcal{E}_{ij}^{+}$ in such a basis.
These operators obey two families of constraints. First we have:
%
\begin{equation}
\mathcal{E}_{ij}^{+}\mathcal{E}_{jk}^{+}=\exp(i2\pi\frac{m}{n^{2}}%
)\mathcal{E}_{jk}^{+}\mathcal{E}_{ij}^{+} \label{Econstraint1}%
\end{equation}
%
Here the sign corresponds to the geometry when link $(jk)$ is perpendicular to
$(ij)$ and located on its left. Notions of left and right are defined for an
observer moving along the link in agreement with its orientation. We have used
the condition $\nu=m/\pi$. Second, these operators are related to the
generators of local gauge transformations by:
%
\begin{equation}
\prod_{j}^{(i)}\mathcal{E}_{ij}^{+}=\exp(i2\pi\frac{m}{n^{2}}\sum_{\mathbf{r}%
}^{(i)}\sigma_{\mathbf{r}})U_{i} \label{Econstraint2}%
\end{equation}
%
As usual, the product in the left-hand side runs over the nearest neighbors of
site $i$, whereas the sum in the right-hand side involves the four plaquettes
adjacent to $i$. In this expression, attention should be paid to the ordering
of the various operators. We assume that the two operators involving bonds
along a given direction directly follow each other. Once this is enforced, any
of the eight possible residual permutations compatible with this criterion
does not change the result. For any oriented link $(ij)$, let us call
$\mathbf{r}$ (resp. $\mathbf{r^{\prime}}$) the adjacent plaquette located at
the left (resp. right) of $(ij)$. The operator $\mathcal{E}_{ij}^{+}$
decreases the local flux $\sigma_{\mathbf{r^{\prime}}}$ and increases
$\sigma_{\mathbf{r}}$ by one unit. We see that $\mathcal{E}_{ij}^{+}$ should
be proportional to $\sigma_{\mathbf{r}}^{+}\sigma_{\mathbf{r^{\prime}}}^{-}$
up to a phase factor which depends on the configuration of local fluxes on the
whole lattice. This phase factor is required in order to satisfy the
constraints~(\ref{Econstraint1},\ref{Econstraint2}) above. In a general case
this phase factor might become a very non-local function of the flux
configuration but it remains simple in the case of
${\mathchoice {\hbox{$\sf\textstyle Z\kern-0.4em Z$}} {\hbox{$\sf\textstyle
Z\kern-0.4em Z$}} {\hbox{$\sf\scriptstyle
Z\kern-0.3em Z$}} {\hbox{$\sf\scriptscriptstyle Z\kern-0.2em Z$}}}_{2}$ $m=2$
model. We discuss now its construction in different cases, starting with the
simplest ones.

We start with the simplest case $n=2$, and \mbox{$m=2$}. Then, the electrical
operators $\mathcal{E}_{ij}^{+}$ on two adjacent and perpendicular links
\emph{anticommute}. Furthermore, according to equation~(\ref{commshiftgauge}),
the generators of the local gauge transformations commute with the shift
operators $S_{jk}=(\widetilde{\pi}_{jk}^{+})^{2}$. It is then possible to
impose \emph{simultaneously} the gauge invariance constraint and the
${\mathchoice {\hbox{$\sf\textstyle Z\kern-0.4em Z$}} {\hbox{$\sf\textstyle
Z\kern-0.4em Z$}} {\hbox{$\sf\scriptstyle
Z\kern-0.3em Z$}} {\hbox{$\sf\scriptscriptstyle Z\kern-0.2em Z$}}}_{2}$
periodicity on the links. As a result, the two operators $\sigma_{\mathbf{r}%
}^{+}$ and $\sigma_{\mathbf{r}}^{-}$ are equal. We may then represent the
above algebra by the following Pauli operators, associated to a quantum Ising
model attached to the plaquettes of the lattice:
%
\begin{align}
\mathcal{E}_{ij}^{+}  &  =\sigma_{\mathbf{r}}^{x}\sigma_{\mathbf{r^{\prime}}%
}^{x}\quad\text{for vertical (ij)}\label{mapping1}\\
&  \mathrm{or}\nonumber\\
\mathcal{E}_{ij}^{+}  &  =\sigma_{\mathbf{r}}^{z}\sigma_{\mathbf{r^{\prime}}%
}^{z}\quad\text{for horizontal (ij)}\label{mapping2}\\
\exp(i\pi\sigma_{\mathbf{r}})  &  =\sigma_{\mathbf{r}}^{y}%
\end{align}
%

The symmetry operators defined in equations~(\ref{constcontP}) and
(\ref{constcontQ}) are easily generalized to the case of a discrete group. We
now have:
%
\begin{align}
\tilde{P}_{\mathrm{r}}  &  =\prod_{j\in\mathrm{row(r)}}\mathcal{E}%
_{j,j+\mathbf{\hat{x}}}^{\mathrm{sign}(j)}\label{constdiscP}\\
\tilde{Q}_{\mathrm{s}}  &  =\prod_{j\in\mathrm{column(s)}}\mathcal{E}%
_{j,j+\mathbf{\hat{y}}}^{\mathrm{sign}(j)} \label{constdiscQ}%
\end{align}
%
In these equations, the symbol $\mathrm{sign}(j)$ stands for $+$ or $-$,
according to the parity of $x_{j}$ (resp. $y_{j}$) in the first (resp. second)
equation, $\mathrm{row(r)}$ denotes row number $\mathrm{r}$ and
$\mathrm{column(s)}$ denotes column number $\mathrm{s.}$ For a generic (not
boundary) row and column these operators commute.

It is now possible to recover the symmetry operations $P_{\mathrm{i}}$ and
$Q_{\mathrm{j}}$ introduced in Section II for Chern-Simons theories with
proper boundary conditions. First, notice that for
${\mathchoice {\hbox{$\sf\textstyle Z\kern-0.4em Z$}} {\hbox{$\sf\textstyle
Z\kern-0.4em Z$}} {\hbox{$\sf\scriptstyle
Z\kern-0.3em Z$}} {\hbox{$\sf\scriptscriptstyle Z\kern-0.2em Z$}}}_{2}$ model
operators $\tilde{P}_{\mathrm{r}}$ and $\tilde{Q}_{\mathrm{s}}$ correspond to
the product of two $P$ (or two $Q$) operators: $\tilde{P}_{\mathrm{r}}%
=P_{r}P_{r+1}$ when expressed in terms of the spin operators of Section II. If
operator $P_{0}=1$ is trivial ($P_{0}=1$), $\tilde{P}_{\mathrm{1}}=P_{1}$ and
we can recover a single $P_{r}$ operator taking the product of all
$P_{r^{\prime}}$ operators with $r^{\prime}<r$. In terms of the gauge theory,
the Chern-Simons Hamiltonian (Eq.~(\ref{HCSZn})) in its $\mu=0$ limit should
not contain the dynamical variables (gauge fields) associated with the bonds
along the edge of the lattice. In this case the operators $\tilde
{P}_{\mathrm{1}}$ and $\tilde{Q}_{\mathrm{1}}$ indeed do not commute because
they contain only one pair of non-commuting electric fields located in the
corner of the lattice. With this assumption, we obtain two families of
operators commuting with the Chern-Simons Hamiltonian at $\mu=0$:
%
\begin{align*}
P_{\mathrm{r}}  &  =\prod_{y_{j}\leq y_{\mathrm{row(r)}}}\mathcal{E}%
_{j,j+\mathbf{\hat{x}}}^{\mathrm{sign}(j)}\\
Q_{\mathrm{s}}  &  =\prod_{x_{j}\leq x_{\mathrm{column(s)}}}\mathcal{E}%
_{j,j+\mathbf{\hat{y}}}^{\mathrm{sign}(j)}%
\end{align*}
%
Here $\mathrm{sign}(j)$ is defined as above. These operators satisfy the
generalized commutation relations:
\begin{equation}
P_{\mathrm{r}}Q_{\mathrm{s}}=\exp(i2\pi\frac{m}{n^{2}})Q_{\mathrm{s}%
}P_{\mathrm{r}}%
\end{equation}
In the special case $m=2$, these operators anticommute. Using
equations~(\ref{mapping1}) and (\ref{mapping2}) above, we may write:
\begin{align*}
P_{\mathrm{r}}  &  =\prod_{\mathbf{r}\in\mathrm{row(r)}}\sigma_{\mathbf{r}%
}^{z}\\
Q_{\mathrm{s}}  &  =\prod_{\mathbf{r}\in\mathrm{column(s)}}\sigma_{\mathbf{r}%
}^{x}%
\end{align*}

We now discuss the mapping in a more general case. First, we assume that the
generators of local gauge transformations still commute with the shift
operators $S_{jk}$ on the links. According to equation~(\ref{commshiftgauge}),
this requires $m$ to be a multiple of $n$. In this case, we may still view the
local fluxes $\sigma_{\mathbf{r}}$ as defined modulo $n$. This implies in
particular that $(\sigma_{\mathbf{r}}^{+})^{n}$ can be chosen to act as the
identity operator. After enforcing the gauge invariance and the
${\mathchoice {\hbox{$\sf\textstyle Z\kern-0.4em Z$}} {\hbox{$\sf\textstyle
Z\kern-0.4em Z$}} {\hbox{$\sf\scriptstyle
Z\kern-0.3em Z$}} {\hbox{$\sf\scriptscriptstyle Z\kern-0.2em Z$}}}_{n}$
periodicity in the bond variables, we may represent $\mathcal{E}_{ij}^{+}$
as:
%
\begin{equation}
\mathcal{E}_{ij}^{+}=\exp(i\mathcal{A}_{\mathbf{r},\mathbf{r^{\prime}}%
}(\{\sigma_{\mathbf{r}}\}))\sigma_{\mathbf{r}}^{+}\sigma_{\mathbf{r^{\prime}}%
}^{-} \label{Eijvertical}%
\end{equation}
%
The \textquotedblright statistical\textquotedblright\ gauge field
$\mathcal{A}_{\mathbf{r},\mathbf{r^{\prime}}}$ should not be confused with the
original link variables $A_{ij}$. This new entity is imposed to us by the
necessity to satisfy the constraints~(\ref{Econstraint1},\ref{Econstraint2}).
Note that as usual, there is a large amount of freedom in the definition of
$\mathcal{A}_{\mathbf{r},\mathbf{r^{\prime}}}$, reflecting the arbitrariness
in choosing a global phase for each state $|\Psi(\{\sigma_{\mathbf{r}%
}\})\rangle$. If we could ignore the first constraint~(\ref{Econstraint1}),
our system of fluxons would be completely equivalent to a collection of
particles obeying fractional statistics, since the second
constraint~(\ref{Econstraint2}) relates the total statistical flux seen by a
fluxon hopping around an elementary plaquette of the dual lattice centered at
$i$, to the number of fluxons in the immediate neighborhoood of $i$. The
presence of the first constraint is an original feature of fluxon dynamics in
Chern-Simons theories.

Let us now show how to construct explicitely one realization for this
statistical gauge field $\mathcal{A}_{\mathbf{r},\mathbf{r^{\prime}}}$. This
amounts to making a definite choice for the global phase of the basis states
$|\Psi(\{\sigma_{\mathbf{r}}\})\rangle$. If we have one fluxon on the
plaquette centered at $\mathbf{r}$, it is possible to represent this by a
string-like configuration of bond variables $p_{ij}(\mathbf{r})$ such that
$p_{ij}$ vanishes on any horizontal bond, and on most vertical bonds, with the
exception of all the links located on the same row as $\mathbf{r}$ and on its
right, for which it takes the value $1$. This may be summarized by the
following definition:
\[
|\mathbf{r}\rangle=\mathcal{P}_{\mathrm{inv}}\prod_{(ij)>\mathbf{r}%
}\mathcal{E}_{ij}^{+}|0\rangle
\]
Here, the operator $\mathcal{P}_{\mathrm{inv}}$ is the projector on the
subspace of the gauge invariant states. The notation $(ij)>\mathbf{r}$ stands
for all the links $(ij)$ on the right-hand side of the plaquette center
$\mathbf{r}$ and on the same row, and the reference state $|0\rangle$ is
simply the state where all the bond variables $p_{ij}$ are equal to $0$. For a
general flux configuration $\{\sigma_{\mathbf{r}}\}$, we simply choose a
reference configuration $p_{ij}(\{\sigma_{\mathbf{r}}\})$ obtained by
superposing the configurations associated to each fluxon excitation in the
system. Therefore, the integers $p_{ij}(\{\sigma_{\mathbf{r}}\})$ are defined
as follows. They are equal to zero for any \emph{horizontal} bond $(ij)$. For
a \emph{vertical} bond, we set:
\[
p_{ij}(\{\sigma_{\mathbf{r}}\})=\sum_{\mathbf{r}<(ij)}\sigma_{\mathbf{r}}
\]
Using this, we define:
%
\begin{equation}
|\Psi(\{\sigma_{\mathbf{r}}\})\rangle=\mathcal{P}_{\mathrm{inv}}\prod
_{(ij)}(\mathcal{E}_{ij}^{+})^{p_{ij}(\{\sigma_{\mathbf{r}}\})}|0\rangle
\label{Definvfluxstates}%
\end{equation}
%

\begin{figure}[th]
\includegraphics[width=3.0in]{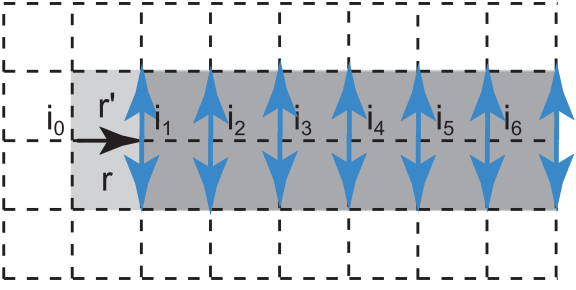}\caption{The electrical field operator,
$\mathcal{E}_{i_{0}i_{1}}$ moves one flux quantum from the plaquette $r$ to
$r^{\prime}$. In order to represent this process in the gauge where all
horizontal links $A_{ij}$ are $0$, we use the gauge transformation on the
string of sites $i_{1}\ldots i_{n}$ that produce the electric field operators
on vertical links (shown by light arrow) and the phase factors associated with
the shaded area. In the flux representation the effect of vertical electric
field is to remove the flux from plaquette $r$ and to add one flux to
plaquette $r^{\prime}$. Note that the phase factors associated with all
plaquettes except $r$ and $r^{\prime}$ are doubled. }%
\label{Tail}%
\end{figure}

We could have used operators $\pi_{ij}^{+}$ instead of $\mathcal{E}_{ij}^{+}$
to generate a state with the same desired flux configuration $\{\sigma
_{\mathbf{r}}\}$, but the advantage of $\mathcal{E}_{ij}^{+}$ is that they
commute with the projector $\mathcal{P}_{\mathrm{inv}}$. These states
$|\Psi(\{\sigma_{\mathbf{r}}\})\rangle$ form a complete basis of the gauge
invariant subspace, since any configuration $\{p_{ij}\}$ of bond variables
producing the flux pattern $\{\sigma_{\mathbf{r}}\}$ may be deduced from the
configuration $\{p_{ij}(\{\sigma_{\mathbf{r}}\})\}$ by a gauge transformation.
With this choice of gauge, we have:
%
\begin{equation}
\mathcal{E}_{ij}^{+}=\sigma^{+}_{\mathbf{r}}\sigma^{-}_{\mathbf{r^{\prime}}}
\label{EijVertical2}%
\end{equation}
for any \emph{vertical} link $(ij)$. This means that $\mathcal{A}%
_{\mathbf{r},\mathbf{r^{\prime}}}$ vanishes whenever the vector joining
$\mathbf{r}$ and $\mathbf{r^{\prime}}$ is equal to $\pm\hat{\mathbf{x}}$. For
a \emph{horizontal} link $(ij)$ oriented from left to right, we make a
repeated use of the second constraint~(\ref{Econstraint2}) above for all the
sites $k$ located on the same row as $(ij)$ and on its right, which, if
applied on a gauge invariant state, yields:
\begin{equation}
\mathcal{E}_{ij}^{+}= \exp\left(  i2\pi\frac{m}{n^{2}} (\sigma_{\mathbf{r}%
}+\sigma_{\mathbf{r^{\prime}}}+2\sum_{\mathbf{r^{\prime\prime}}>\mathbf{r}%
,\mathbf{r^{\prime}}} \sigma_{\mathbf{r^{\prime\prime}}})\right)
\mathcal{M}_{ij}\nonumber
\end{equation}
where the ''string'' operator $\mathcal{M}_{ij}$ is defined by:
\[
\mathcal{M}_{ij}=\prod_{(kl)>\mathbf{r}}\mathcal{E}_{kl}^{+} \prod
_{(kl)>\mathbf{r^{\prime}}}\mathcal{E}_{kl}^{-}
\]
As before, $\mathbf{r}$ (resp. $\mathbf{r^{\prime}}$) denotes the adjacent
plaquette located above (resp. below) the oriented link $(ij)$. The notation
$(kl)>\mathbf{r}$ refers to all the vertical links $(kl)$ on the right of
plaquette $\mathbf{r}$ and on the same row. In our gauge the string operator
becomes
\[
\mathcal{M}_{ij}|\{\sigma_{\mathbf{r}}\}\rangle= \sigma^{+}_{\mathbf{r}}%
\sigma^{-}_{\mathbf{r^{\prime}}} |\{\sigma_{\mathbf{r}}\}\rangle
\]

Finally, we get
%
\begin{equation}
\mathcal{E}_{ij}^{+}=\sigma^{+}_{\mathbf{r}}\sigma^{-}_{\mathbf{r^{\prime}}}
\exp\left(  i2\pi\frac{m}{n^{2}} (\sigma_{\mathbf{r}}+\sigma
_{\mathbf{r^{\prime}}}+2\sum_{\mathbf{r^{\prime\prime}}>\mathbf{r}%
,\mathbf{r^{\prime}}} \sigma_{\mathbf{r^{\prime\prime}}})\right)
\label{Eijhorizontal}%
\end{equation}
where the formula applies only on the gauge invariant subspace. This shows
that
\[
\exp(i\mathcal{A}_{\mathbf{r},\mathbf{r^{\prime}}})=\exp\left(  i2\pi\frac
{m}{n^{2}} (\sigma_{\mathbf{r}}+\sigma_{\mathbf{r^{\prime}}}+2\sum
_{\mathbf{r^{\prime\prime}}>\mathbf{r},\mathbf{r^{\prime}}} \sigma
_{\mathbf{r^{\prime\prime}}})\right)
\]
whenever the vector joining $\mathbf{r}$ and $\mathbf{r^{\prime}}$ is equal to
$\pm\hat{\mathbf{y}}$. The right-hand side of this expression is most of the
time highly \emph{non-local}. The only exception is the case when $2m$ is an
integer times $n^{2}$. For $m$ chosen in the interval \mbox{$[0,n^{2}-1]$},
this occurs only when $2m=n^{2}$, which requires $n$ to be even. If this
condition is satisfied, we get:
\[
\exp(i\mathcal{A}_{\mathbf{r},\mathbf{r}+\mathbf{\hat{y}}}) = \exp(i\pi
(\sigma_{\mathbf{r}}+\sigma_{\mathbf{r^{\prime}}}))
\]
In physical terms, this set of models corresponds to an effective Bose
statistics for fluxons.

Let us now address the most general case. Since according
to~(\ref{commshiftgauge}), exchanging $U_{i}$ and $S_{jk}$ involves a phase
factor $\exp(\pm i2\pi\frac{m}{n})$, we may always diagonalize simultaneously
the local gauge generators $U_{i}$ with the operators $S_{jk}^{n^{\prime}}$
where $n^{\prime}$ is equal to $n$ divided by the greatest common divisor of
$m$ and $n$. This implies that dimension of the Hilbert space associated with
one flux is increased to $N=nn^{\prime}$ and we may view the local fluxes as
elements of
${\mathchoice {\hbox{$\sf\textstyle Z\kern-0.4em Z$}}{\hbox{$\sf\textstyle
Z\kern-0.4em Z$}}{\hbox{$\sf\scriptstyle
Z\kern-0.3em Z$}}{\hbox{$\sf\scriptscriptstyle Z\kern-0.2em Z$}}}_{N}$. All
the expressions already written are then valid. In particular, we may check
that $\exp(i\mathcal{A}_{\mathbf{r},\mathbf{r^{\prime}}})$ is unchanged, if we
change $\sigma_{\mathbf{r}}$ into $\sigma_{\mathbf{r}}+Nb_{\mathbf{r}}$, where
$b_{\mathbf{r}}$ is any integer.

The representation in terms of fluxes constructed in this Section becomes
especially convenient when the gap of a single fluxon is the largest energy
scale in the problem which occurs if $\mu>>\lambda_{n}^{-1}$ (see
Hamiltonian~(\ref{HCSZn})). In this limit, the term proportional to
$\lambda_{n}^{-1}$ induces tunneling processes where a fluxon jumps from a
plaquette to one of its neighbors. Using the constraints~(\ref{Econstraint1}%
)(\ref{Econstraint2}), we see that the operator which moves a fluxon around
the elementary plaquette of the dual lattice centered at site $i$ is simply
equal to $U_{i}$, provided that it acts on a state with only one fluxon on a
plaquette adjacent to $i$. Indeed, this operator may be written as:
$\mathcal{E}_{i,i-\mathbf{\hat{y}}}^{+}\mathcal{E}_{i,i-\mathbf{\hat{x}}}%
^{+}\mathcal{E}_{i,i+\mathbf{\hat{y}}}^{+}\mathcal{E}_{i,i+\mathbf{\hat{x}}%
}^{+}$; notice that this ordering is different from the
equation~(\ref{Econstraint2}). To recast this product in the form of the
latter expression, one may permute the last two operators on the right, which
according to~(\ref{Econstraint1}) produces a phase factor $\exp(-i2\pi\frac
{m}{n^{2}})$. But for a state with exactly one fluxon located next to site
$i$, this factor is exactly cancelled by the term $\exp(i2\pi\frac{m}{n^{2}%
}\sum_{\mathbf{r}}^{(i)}\sigma_{\mathbf{r}})$ present in
equation~(\ref{Econstraint2}), leaving only the local gauge generator $U_{i}$.
This shows that a single fluxon moves in this limit as a free quantum particle
on the dual lattice, with an energy spectrum:
\[
\epsilon(k)=\mu(1-\cos(\frac{2\pi}{n}))-2\lambda_{n}^{-1}(\cos(k_{x}%
)+\cos(k_{y}))
\]
This spectrum is gapped as long as $\lambda$ remains larger than $\lambda
_{n}^{c}$ given by:
\begin{equation}
\frac{1}{\lambda_{n}^{c}}=\frac{\mu}{4}(1-\cos(\frac{2\pi}{n}))
\label{lambda^c_n}%
\end{equation}
These equations neglects the renormalization of the fluxon spectrum by the
fluctuations, but we expect that it remains qualitatively correct when these
fluctuations are taken into account. In particular, even in the regime of the
strong fluctuations, close to the transition, the picture of the fluxons
moving with a spectrum $\epsilon(k)=\delta\epsilon+\frac{1}{2m}\mathbf{k}^{2}$
should remain valid at sufficiently long scales as long as $\lambda
_{n}>\lambda_{n}^{c}$. For a system with an external charge $Q_{i}$ located at
site $i$, we have $U_{i}=\exp(i\frac{2\pi}{n}Q_{i})$, so a single fluxon will
experience the usual Aharonov-Casher effect from this static charge,
independently of the value of the Chern-Simons coefficient $\nu$. Since this
interference effect always raises the value of the fluxon energy, in
comparison to the case $Q_{i}=0$, we see that the flux attachment mechanism
predicted in the continuous $U(1)$ Chern-Simons theory does not operate as
long as the single fluxon spectrum remains gapped. In the notations of this
section~(cf. (\ref{CS}) above), this implies that $\lambda_{n}^{\ast}%
=\lambda_{n}^{c}$.

\section{Conclusion}

We presented (Section II)\ the general symmetry analysis of physical systems
with protected degeneracies, i.e. with degeneracies that are exponentially
weakly affected by local perturbations. We have shown that such protected
degeneracies appear in a system described by a wide class of Hamiltonians that
commute with two sets of integrals of motion, $\{P_{i}\}$ and $\{Q_{j}\} $ but
which do not commute themselves. These sets of non-commuting operators should
allow for a finite dimensional representation. In the simplest case, if
$P_{i}P_{j}$ and $Q_{k}Q_{l}$ commute with all other operators the algebra of
these operators allows two dimensional representation and the states of the
system are exactly doubly degenerate. For the effect of local perturbations to
become really small in the thermodynamic limit one also needs that the gap to
the low energy excitations remains finite.

We have explicitly constructed a two dimensional lattice spin model with local
interactions and which has these integrals of motion. In this model all states
are exactly doubly degenerate. The behavior of this spin model is
characterized by the dimensionless parameter, $J_{z}/J_{x}$ which physically
corresponds to the anisotropy of the couplings in different directions. We
were able to treat it analytically in the regime of large (or small)\ values
of this parameter. In this regime the spectrum of the system contains $2^{L}$
low energy modes where $L$ is the linear size of the spin array. The gap
between these modes and the ground state decreases exponentially with the
system size. The number of these low energy modes is the same as would be the
number of edge states but, unlike the latter, they are not sensitive to the
boundary conditions. In order to check the validity of these conclusions for
all values of $J_{z}/J_{x}$ we have also performed the diagonalization of
small arrays (up to 25 spins) and concluded that the gap to low energy states
remains a decreasing function of the system size for all values but this
decrease becomes very slow for $J_{z}/J_{x}\sim1$. It remains unclear to us,
however, whether the system exhibits a new phase at these values of the
parameter or this apparently slower decrease of the gap is a consequence of a
critical behaviour.

We have suggested and studied (Section III)\ two designs of the Josephson
junction arrays and showed that their effective low energy Hamiltonians
satisfy the symmetry requirements described above and thus their states are
doubly degenerate and protected from the external noise. The simplest of these
arrays can be mapped onto a spin model with non-local interactions. The
non-locality of these interactions, however, is not important for the
protection from the external noise. Further, in these systems one can
completely eliminate the dangerous low energy modes by appropriate boundary
conditions. The mapping of the Josephson junction array onto a spin system
with symmetric Hamiltonian implies that the continuous superconducting phase
can be integrated out. We have examined the conditions when this can be done
and when low energy degrees of freedom corresponding to the continuous phase
are irrelevant. Summarizing the requirements for a physical Josephson
junctions we conclude that they are relatively easy to satisfy in medium sized
arrays (up to $10\times10$ elements) which should be quite sufficient to get a
noise suppression by 10 orders of magnitude.

The spin models studied in Section II\ can be also mapped onto a discrete
Chern-Simons theory on a lattice. In order to establish this mapping we have
constructed (Section IV)\ a Hamiltonian framework for lattice Chern-Simons
theories with abelian groups. We argued that, in contrast to the continuous
theories, such theories generally have low energy modes corresponding to the
excitations with large momentum, comparable to the inverse lattice spacing.
Further, we showed that in a theory with a compact group (in particular, in a
theory with discrete group) the Chern-Simons coefficient, $\nu$, is quantized
similar to the quantization of the magnetic flux through the torus: $\nu
=\pi/m$. In the gauge invariant space of magnetic fluxes the kinetic part of
the Hamiltonian of these theories can be described as a flux dynamics. Due to
the presence of Chern-Simons term the motion of fluxes in $x $ and $z$
directions does not commute. In the simplest case of the
${\mathchoice {\hbox{$\sf\textstyle Z\kern-0.4em Z$}}{\hbox{$\sf\textstyle
Z\kern-0.4em Z$}}{\hbox{$\sf\scriptstyle
Z\kern-0.3em Z$}}{\hbox{$\sf\scriptscriptstyle Z\kern-0.2em Z$}}}_{2}$, $m=2$
theory, the fluxes take only two values and their motions in $x$ and $z$
directions anticommute allowing us to map this theory onto the spin model
studied in Section II. In a general case the Hamiltonian in flux
representation becomes very non-local but still this representation is
convenient in the limit of large magnetic energy when the fluxes are rare.
Using this limit, we show that the flux attachment to the charge only occurs
if a single fluxon is gapless.

The main theoretical issue raised by the Sections IV and V of the paper is the
precise connection between lattice and continuous versions of Chern-Simons
theories. For the continuous case, there is a sharply isolated ground-state
subspace, whose degeneracy directly reflects the topology of the
two-dimensional space on which the model is defined. In a recent series of
papers~\cite{Freedman2003a,Freedman2003b}, various descriptions of these
models (in terms of wave-functions defined on equivalence classes of loops)
have been advocated to construct candidate lattice models which would exhibit
a ground-state sector equivalent to a pure (topological) Chern-Simons theory.
The approach we have followed here starts from a direct quantization of a
lattice Hamiltonian inspired from the continuous Chern-Simons theory with an
additional kinetic term. We found that such construction typically leads to
the degenerate modes attached to the Brillouin zone boundary; it remains to be
investigated whether or not the presence of these modes spoils the expected
properties of continuous models (such as statistical transmutation of external
charges). It would be also interesting to see whether these degeneracies
remain for other lattice structures. Finally, extending this construction to
non-Abelian discrete groups is clearly desirable, from the perspective of
enlarging the set of unitary operations generated by adiabatic exchanges
between charge and/or vortex
excitations~\cite{Kitaev1997,Mochon2003a,Mochon2003b}.

\textbf{Acknowledgements}

We are thankful to L. Faoro, A. Kitaev and J. Vidal for useful discussions.
LI\ is thankful to LPTHE, Jussieu for their hospitality while MF and BD have
enjoyed the kind hospitality of the Physics Department at Rutgers University.
This work was made possible by support from NATO CLG grant 979979, NSF DMR
0210575, RFBR grant 04-02-16348-a. and by CNRS, Russian Academy of Science
under Program "Quantum Macrophysics" \ and Russian Ministry of Science under
Project "Physics of Quantum Computing".

\appendix

\section{Exact solution of two chain problem}

For the diagonalization of the hamiltonian (\ref{H_col}) it is convenient to
rotate the Pauli matrices: $\tau_{i}^{z}\rightarrow\tilde{\sigma}_{i}%
^{x},\;\;\tau_{i}^{x}\rightarrow\tilde{\sigma}_{i}^{z}$ and use the
Jordan-Wigner fermionic representation (see \cite{Fradkin})
\begin{align}
\tilde{\sigma}_{i}^{+}  &  =a_{i}^{+}\exp\left\{  i\pi\sum_{i^{\prime}%
=1}^{i-1}a_{i^{\prime}}^{+}a_{i^{\prime}}\right\}  ,\nonumber\\
\tilde{\sigma}_{i}^{-}  &  =\exp\left\{  -i\pi\sum_{i^{\prime}=1}%
^{i-1}a_{i^{\prime}}^{+}a_{i^{\prime}}\right\}  a_{i},\nonumber\\
\tilde{\sigma}_{i}^{z}  &  =2a_{i}^{+}a_{i}-1. \label{j-w}%
\end{align}
so that (\ref{H_col}) takes the form
\begin{align}
H  &  =2J_{z}\left\{  \sum_{l=1}^{n-1}(a_{l}^{+}-a_{l})(a_{l+1}+a_{l+1}%
^{+})\right. \nonumber\\
&  \left.  +2\Lambda\sum_{l=1}^{n}(a_{l}^{+}a_{l}-1/2)\right\}  , \label{ham6}%
\end{align}
where $\Lambda=J_{x}/2J_{z}$. By means of the Bogoliubov-like transformation
the hamiltonian (\ref{ham6}) can be diagonalized
\[
H=\sum_{k}E_{k}(2b_{k}^{+}b_{k}-1)
\]
where $b_{k}^{+},b_{k}$ are the fermionic operators of the eigenmodes with
eigenenergies
\[
E(k)=2J_{z}\sqrt{\Lambda^{2}+1+2\Lambda\cos k}.
\]

The quasi-continuous spectrum of this hamiltonian can be found from the
quantization rule for the quasimomentum $k$:
\begin{align}
k(n+1)-\arctan\left(  \frac{\sin k}{\Lambda+\cos k}\right)  =\pi m,
\label{solution2}%
\end{align}
with integer $m$. For $\Lambda>1$ this equation has exactly $n$ distinct
nontrivial solutions, and the set of the corresponding eigenfunctions is
complete. It is not the case for $\Lambda<1$, however. Here the number of the
continuous spectrum eigenstates is only $n-1$, so that there should be one
additional mode - the bound state. To find the latter one should look for
complex solutions of the dispersion equation (\ref{solution2}). Substituting
$k=\pi+i\gamma$, we arrive at
\begin{align}
\gamma(n+1)=\frac12\ln\left\{  \frac{\Lambda-e^{\gamma}}{\Lambda-e^{-\gamma}%
}\right\}  , \label{solution11l}%
\end{align}
and, introducing $e^{-\gamma}=\Lambda+x$ with small $x\ll1$, we get
$x\approx\left(  \frac{1}{\Lambda}-\Lambda\right)  \Lambda^{2(n+1)}$, so that
and the bound state energy is
\begin{align}
2\Delta=2E(\pi+i\gamma) \approx4J_{z}[1-(J_{x}/2J_{z})^{2}](J_{x}/2J_{z})^{n}.
\label{solution10r}%
\end{align}
The formula (\ref{solution10r}) is valid for $(J_{x}/2J_{z})^{n}\ll1$. In the
fermionic representation the bound-state eigenfunction is localized near the
ends of the chain within the range $\xi\sim-\ln\Lambda$. The localized
character of the mode responsible for the ground state doublet splitting is,
however, an artifact of the nonlocal representation (\ref{j-w}) and,
apparently, does not have much physical meaning.

A similar result for the splitting can also be obtained for an Ising chain
with periodic boundary conditions. Here the chain does not have ends, and the
dispersion equation has only solutions, corresponding to the continuous
spectrum. The splitting arises from the following effect: the effective
boundary conditions for the Jordan-Wigner fermions $a$ are periodic or
antiperiodic, depending on the parity $I$ of their total number (the latter is
a good quantum number):
\begin{align}
kn=\left\{
\begin{aligned} 2\pi m, & \qquad\mbox{for $I=1$},\\ 2\pi (m+1/2), & \qquad\mbox{for $I=-1$}, \end{aligned} \right.
\label{solution2g}%
\end{align}
with $m=0,1,\ldots,L-1$.

The energy of the first excited state is just due to this effect; it comes not
from any specific single-particle state, but from the entire Fermi sea of the
filled energy levels, each of which is slightly shifted when the parity $I$ is
changing:
\begin{align}
2\Delta\! = \! \sum_{m=0}^{n-1} \! \left\{  E\left(  k \! = \! \frac{2\pi m}
{n}\right)  \! - \! E\left(  k \! = \! \frac{2\pi(m+1/2)}{n}\right)  \right\}
\label{solution5y}%
\end{align}
Using the Poisson summation formula we arrive at the result
\begin{align}
2\Delta\approx4J_{z}\sqrt{ \frac{1-(J_{x}/2J_{z})^{2}}{\pi n}}(J_{x}%
/2J_{z})^{n}, \label{solution7y}%
\end{align}
which differs from (\ref{solution10r}) only in the preexponential factor.

\section{The critical point of three chain problem}

We consider a three-strings ladder with periodic boundary conditions along
each rung. The corresponding Hamiltonian reads
\begin{equation}
H=-J_{z}\sum_{i=1}^{n-1}\sum_{j=1}^{3}\sigma_{ij}^{z}\sigma_{i+1j}^{z}%
-J_{x}\sum_{i=1}^{n}\sum_{j=1}^{3}\sigma_{ij}^{x}\sigma_{ij+1}^{x},
\label{ham3}%
\end{equation}
where $j=4$ is identical to $j=1$. We introduce a basis of 4 states $\psi_{m}$
(with $m=0,1,2,3$) on a particular rung $i$, corresponding to the sector with
all $P_{i}=1$:
\[
\psi_{0}=\left(
\begin{aligned} \uparrow\\\uparrow\\\uparrow \end{aligned}\right)  ,\quad\!
\psi_{1}=\left(  \begin{aligned}
\uparrow\\\downarrow\\\downarrow\end{aligned}\right)  , \quad\! \psi
_{2}=\left(
\begin{aligned} \downarrow\\\uparrow\\\downarrow\end{aligned}\right)  ,
\quad\! \psi_{3}=\left(  \begin{aligned}
\downarrow\\\downarrow\\\uparrow\end{aligned}\right)  ,
\]
Then (up to an irrelevant additive constant) the Hamiltonian (\ref{ham3}) can
be rewritten as
\[
H=-4J_{z}\sum_{i=1}^{n-1}\delta_{m_{i}m_{i+1}}-J_{x}\sum_{i=1}^{n}\Gamma_{i},
\]
where the matrix $\Gamma$
\[
\hat{\Gamma}=\left(  \begin{aligned} \quad 0\quad & \quad 1\quad &\quad
1\quad &\quad 1\quad \\ \quad 1\quad& \quad 0\quad &\quad 1\quad &\quad
1\quad \\ \quad 1\quad & \quad 1\quad &\quad 0\quad &\quad 1\quad \\ \quad
1\quad & \quad 1\quad &\quad 1\quad &\quad 0\quad \end{aligned}\right)  ,
\]
plays a role, similar to that of the $\sigma_{x}$-operator for the two chain
problem. This is the hamiltonian of the one dimensional $q=4$ Potts model in a
"transverse field".

Consider now an asymmetric classic two-dimensional $q=4$ Potts model with the
Hamiltonian
\begin{align}
\beta H_{2d}=-K_{z}\sum_{ik}\delta_{m_{ik}m_{i+1k}}- K_{x}\sum_{ik}%
\delta_{m_{ik}m_{ik+1}} , \label{part1}%
\end{align}
The transfer-matrix for this system (in the $k$-direction) is
\begin{align}
\hat{T}=\exp\left\{  K_{z}\sum_{i}^{n-1}\delta_{m_{i}m_{i+1}}\right\}
\prod_{i=1}^{n}\left(  \hat{1}e^{K_{x}}+\hat{\Gamma}_{i}\right)  ,
\label{part2}%
\end{align}
where $\hat{1}$ is $4\times4$ unity matrix. Using the identity
\[
e^{h\hat{\Gamma}}= \frac14\left\{  (e^{3h}+3e^{-h})+(e^{3h}-e^{-h})\hat
{\Gamma}\right\}
\]
for matrix $\Gamma$, we can rewrite (\ref{part2}) in the form
\begin{align}
\hat{T}=C\exp\left\{  K_{z}\sum_{i}^{n-1}\delta_{m_{i}m_{i+1}} +h\sum_{i}%
^{n}\Gamma_{i}\right\}  , \label{part2r}%
\end{align}
where $C$ is an irrelevant constant and $h$ is determined by $e^{K_{x}%
}=(e^{4h}+3)/(e^{4h}-1)$.

The line of critical points for the asymmetric two-dimensional $q=4$ Potts
model is governed by the relation (see \cite{Baxter})
\[
(e^{K_{x}}-1)(e^{K_{z}}-1)=4;
\]
in terms of $K_{z},h$ this relation takes simple form $K_{z}=4h$. On the other
hand, the matrix (\ref{part2r}) describes the time evolution of the quantum
system with the Hamiltonian (\ref{ham3}) and with $4J_{z}=\frac{K_{z}}{\Delta
t},\quad J_{x}=\frac{h}{\Delta t}$, where $\Delta t\rightarrow0$ is an
infinitesimal time-interval. Thus, we conclude, that the quantum phase
transition in our initial three chain system takes place at the symmetric
point $J_{x}=J_{z}$.

Unfortunately, the solution of the two-dimensional $q=4$ Potts model away from
the critical line is not known and, in contrast to exactly solvable two-chain
model, we can not find the dependence of the gap $\Delta$ on the parameter
$J_{x}/J_{z}$ in the full range of this parameter.

\end{document}